\documentclass[12pt,a4paper]{article}

\usepackage[margin=2.50cm,nohead]{geometry}
\usepackage{graphics}
\usepackage{epsfig}
\usepackage{amssymb}
\usepackage{dsfont}
\usepackage{color}
\usepackage{cite}
\usepackage{epsf}
\usepackage{axodraw}

\def\theequation{\arabic{section}.\arabic{equation}}

\definecolor{green3}{rgb}{0.,0.7,0.0}
\definecolor{red1}{rgb}{0.9,0,0}

\newcommand{\sla}{\hspace{0.1cm}{\large/}\hspace{-0.25cm}}

\newcommand{\slashed}[1]{\not\!#1}

\def\lsim{\raise0.3ex\hbox{$\;<$\kern-0.75em\raise-1.1ex\hbox{$\sim\;$}}}
\def\gsim{\raise0.3ex\hbox{$\;>$\kern-0.75em\raise-1.1ex\hbox{$\sim\;$}}}
\DeclareMathAlphabet{\scr}{U}{rsfs}{m}{n}

\newcommand{\One}{\mathds{1}}

\newcommand{\CL}{C^{\alpha,L}_{i\tilde j}}
\newcommand{\CR}{C^{\alpha,R}_{i\tilde j}}
\newcommand{\CLs}{C^{\alpha,L\ast}_{i\tilde j}}
\newcommand{\CRs}{C^{\alpha,R\ast}_{i\tilde j}}
\newcommand{\CLb}{C^{\beta,L}_{i\tilde j}}
\newcommand{\CRb}{C^{\beta,R}_{i\tilde j}}

\begin{document}

\begin{flushright}
CERN-PH-TH/2011-165, FTUV-11-1708, MAN/HEP/2011/12\\
arXiv:1108.3314~[hep-ph]\\
\end{flushright}

\begin{center}
{\Large\bf CP Violation in Correlated Production and Decay \\[4mm] 
  of Unstable Particles }
\end{center}

\vspace{3mm}

\begin{center}
{\sc 
Olaf Kittel$^{a,\, b,\, c,\,}$\footnote{Email: kittel@th.physik.uni-bonn.de} and 
Apostolos Pilaftsis$^{b,\, c,\, d,\,}$\footnote{Email:
  apostolos.pilaftsis@manchester.ac.uk} 
}
\end{center}

\begin{center}
$^a${\it Departamento de F\'isica Te\'orica y del Cosmos and CAFPE,
Universidad de Granada,\\ E-18071 Granada, Spain}\\[2mm] 
$^b${\it 
Consortium for Fundamental Physics, School of Physics and Astronomy,\\
University of Manchester, Manchester M13 9PL, United Kingdom}\\[2mm]
$^c${\it Theory Division, CERN, CH-1211
    Geneva 23, Switzerland}\\[2mm]
$^d${\it Department of Theoretical Physics and
    IFIC, University of Valencia-CSIC,\\
E-46100, Valencia, Spain}
\end{center}

\vspace{12mm}

\centerline{\bf ABSTRACT}

\noindent
We study  resonant CP-violating Einstein--Podolsky--Rosen correlations
that may  take place  in the production  and decay of  unstable scalar
particles at high-energy colliders.  We  show that as a consequence of
unitarity  and CPT  invariance  of the  ${\rm  S}$-matrix, in  $2\to2$
scatterings  mediated  by  mixed  scalar  particles,  at  least  three
linearly  independent  decay  matrices  associated with  the  unstable
scalar states  are needed to  obtain non-zero CP-odd  observables that
are  also  odd  under  C-conjugation.   Instead,  for  the  correlated
production  and decay  of  two unstable  particle  systems in  $2\to4$
processes,  we  find that  only  two  independent  decay matrices  are
sufficient to induce a  net non-vanishing CP-violating phenomenon.  As
an application of  this theorem, we present numerical  estimates of CP
asymmetries for the correlated  production and decay of supersymmetric
scalar  top--anti-top pairs  at the  LHC, and  demonstrate  that these
could  reach values of  order one.   As a  byproduct of  our analysis,
we~develop  a novel  spinorial trace  technique, which  enables  us to
efficiently  evaluate   lengthy  expressions  of   squared  amplitudes
describing the resonant scalar transitions.

\bigskip


\newpage

\section{Introduction}\label{sec:Introduction}

In  1935, Einstein,  Podolsky  and Rosen  (EPR)~\cite{EPR} proposed  a
remarkable  experiment,  by~which  the  non-local  nature  of  quantum
mechanics  could  be tested  in  unstable  systems  decaying into  two
entangled   states.   Contrary   to  authors'   initial  expectations,
subsequent EPR-type experiments based  on polarization of photons have
refuted  local  realism,  vindicating  the  non-local  and  non-causal
interpretation  of  the  quantum-mechanical wave\-function.   The  EPR
paradigm   has   found  numerous   applications   in  modern   quantum
theory~\cite{Bell},  such  as quantum  information,  quantum dots  and
particle  physics,  including  tests  of  CPT  violation  and  quantum
decoherence in  $\phi$-factories as  predicted in certain  theories of
quantum gravity~\cite{Bernabeu:2003ym}.

In this paper, we study  the physical consequences of EPR entanglement
on resonant CP violation~\cite{resScalar} in the correlated production
and decay of unstable particles  at high-energy colliders, such as the
TEVATRON and  the CERN $pp$  Large Hadron Collider (LHC)  (for related
considerations   at  future   colliders,  see~\cite{stauCP}).    As  a
prototype example,  we consider the pair  production of SU(3)-coloured
scalar particles $\widetilde{S}_{1,2}$ via  virtual gluons at the LHC,
followed    by    their   decay    into    fermions,   e.g.~$pp    \to
\widetilde{S}_{1,2}  \,   \widetilde{S}_{1,2}^*  \to  (f\,\tilde{f})\:
(\bar{f}'\,\bar{\tilde{f}'})$.    For   instance,  in   supersymmetric
theories, such coloured scalar particles could be scalar top or bottom
quarks,           i.e.~$\widetilde{S}_{1,2}           \in           \{
\tilde{t}_{1,2}\,,\,\tilde{b}_{1,2}\}$,  which   can  decay  into  the
gauginos  or Higgsinos  $\tilde{f}$,  $\bar{\tilde{f}'}$ and  Standard
Model   (SM)  fermions   $f$,   $\bar{f}'$.   As   a  consequence   of
cross-correlation between  unitarity and CPT invariance,  we find that
the    two    decaying    `arms'    of    $\widetilde{S}_{1,2}$    and
$\widetilde{S}^*_{1,2}$ are not independent  of each other. As we will
show, at least  two linearly independent decay channels  are needed to
obtain  non-zero CP  asymmetries.  The  proof of  this  statement gets
facilitated with the use of  a new trace technique, which we introduce
in order to efficiently evaluate lengthy scalar amplitude expressions.

The paper is  organized as follows. After this  brief introduction, in
Section~\ref{sec:formalism} we review  the formalism for describing $2
\to 2$ resonant  scatterings, mediated by a system  of mixed, unstable
scalar particles, and  present the action of CP  and T transformations
on the squared amplitudes.  In Section~\ref{CPTobservables} we discuss
the  unitarity  and   CPT  constraints  on  CP-violating  observables.
Section~\ref{sec:EPR} shows the connection  between $2\to 2$ and $2\to
4$  scatterings, and  presents numerical  estimates of  CP asymmetries
that   may  occur   in  the   correlated  production   and   decay  of
supersymmetric  scalar   top--anti-top  pairs.   In   particular,  the
scattering processes are identified for optimally testing CP violation
in the  scalar top sector  of the Minimal Supersymmetric  extension of
the  Standard  Model   (MSSM)  at  the  LHC.   Section~\ref{sec:concl}
contains  our conclusions.  In  Appendix~\ref{sec:Matrixformalism}, we
give technical details  of a novel spinorial trace  technique which we
develop  to analytically  evaluate lengthy  expressions  of amplitudes
squared describing transitions mediated by unstable scalar particles.

\setcounter{equation}{0}
\section{Formalism for Scalar Particle Mixing}
\label{sec:formalism}

In  this section,  we will  first review  and then  extend  the former
formalism for unstable particle mixing, presented in~\cite{resScalar}.
Our aim  is to  study CP  asymmetries in the  production and  decay of
scalar  particles.   Motivated  by  supersymmetry, we  consider  as  a
prototype  example coloured  scalar  particles $\widetilde{S}_\alpha$,
such  as scalar top  ($\tilde t_{1,2}$)  or bottom  ($\tilde b_{1,2}$)
quarks. We~assume  that these particles are sufficiently  heavy, so as
to  decay  into   two  fermions,  e.g.~$\widetilde{S}_\alpha  \to  f_i
\tilde{f}_{\tilde   j}$,  where   $f_i   \in  \{   t,   \,b  \}$   and
$\tilde{f}_{\tilde  j}\in \{  \tilde  g, \,  \tilde\chi_{1,2,3,4}^0,\,
\tilde\chi_{1,2}^+\}$.  The generic Lagrangian describing the relevant
scalar-fermion-gaugino/Higgsino                            interactions
$\widetilde{S}_\alpha$-$f_i$-$\tilde f_{\tilde j}$ reads:
\begin{eqnarray}
{\scr L}_{ \tilde S_\alpha f_i \tilde f_{\tilde j}}
&=& \widetilde{S}_\alpha\,\bar f_i\,\left(\CL \, P_L + \CR\,P_R\right)
            \,\tilde f_{\tilde j}\  +\ {\rm H.c.}\; ,
\end{eqnarray}
with $P_{L,  R}=(1 \mp \gamma_5)/2$. Note that  explicit couplings for
$\widetilde{S}_\alpha=\tilde   t_\alpha,  \tilde   b_\alpha$   in  the
CP-violating MSSM may be found, for example, 
in~\cite{Bartl:2003he,Eberl:2009xe}.

\begin{figure}[t]
\scalebox{1}{
\begin{picture}(10,5)(-5,0)
	\put( -1.5,2.733){$ i\, \Pi_{\alpha\beta} \;  \; =$}
	\DashArrowLine(10,80)(50,80){5}
	\ArrowLine(25,71)(40,71)
        \put( 1.,2.1){ $p$}
	\Vertex(50,80){1.5}
	   \ArrowArc(70,80)(20,30,210)
	   \ArrowArc(70,80)(20,210,40)
	\Vertex(90,80){1.5}
	\DashArrowLine(90,80)(130,80){5}
	\ArrowLine(105,71)(120,71)
        \put( 3.8,2.1){ $p$}
	\put( 0.8, 3.1){ $\widetilde S_\beta $}
	\put( 3.7, 3.1){ $\widetilde S_\alpha $}
	\put( 1.65, 3.75){ $\tilde f_{\tilde j}, \quad  k-p$}
        \put( 2.7,1.7){ $f_{i}, k $}
%
%
	\DashLine(70,40)(70,120){5}
%
	\Line(72,48)(75,50)
	\Line(72,52)(75,54)
	\Line(72,56)(75,58)
	\Line(72,60)(75,62)
	\Line(72,64)(75,66)
	\Line(72,68)(75,70)
	\Line(72,72)(75,74)
	\Line(72,76)(75,78)
        \Line(72,80)(75,82)
        \Line(72,84)(75,86)
        \Line(72,88)(75,90)
        \Line(72,92)(75,94)
        \Line(72,96)(75,98)
        \Line(72,100)(75,102)
        \Line(72,104)(75,106)
        \Line(72,108)(75,110)

%
\end{picture}}
\caption{\it Scalar  transitions $\widetilde S_\beta  \to \widetilde S_\alpha$
  at the one-loop level.  A cut on the graph according to the Cutkosky
  rules~\cite{CutRules} yields the absorptive contributions.  }
\label{Fig:SiSj}
\end{figure}
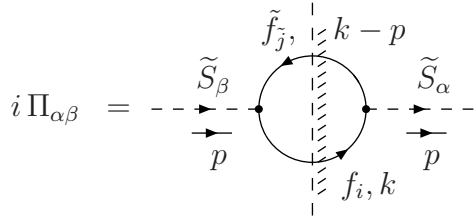

The dynamics  of the  mixed $\widetilde S_{1}\widetilde  S_{2}$ system  may be
described   by   the   (unrenormalized)  inverse   scalar   propagator
matrix~\cite{resScalar}
\begin{eqnarray}
\label{Delta}
 {\Delta}^{-1}_0 (s) &=& 
\left(
 \begin{array}{cc}
 s\: -\: M_1^2\: +\:  \Pi_{11}(s) &  \Pi_{12}(s) \\[2mm]
    \Pi_{21}(s)     & s\: -\: M_2^2\:  +\:  \Pi_{22}(s) 
 \end{array} \right)\; .
\end{eqnarray}
The propagator matrix arises from summing up a geometric series of the
$\widetilde      S_{\alpha}      \widetilde      S_{\beta}$      self-energies
$\Pi_{\alpha\beta}(s)$,   as   shown   in   Fig.~\ref{Fig:SiSj}.   The
self-energies  may  be linearly  decomposed  in  their dispersive  and
absorptive parts as follows:
\begin{eqnarray}
  \label{selfenergies}
\Pi_{\alpha\beta}(s) &=&     \Pi_{\alpha\beta}^{\rm disp}(s)\: 
                        +\: i\, \Pi_{\alpha\beta}^{\rm abs}(s)\; .
\end{eqnarray}
The  dispersive  part   of  the  self-energy,  $\Pi_{\alpha\beta}^{\rm
  disp}(s)$, is UV infinite and requires renormalization. To this end,
  we consider the on-shell (OS)  scheme, in which the dispersive parts
  satisfy the renormalization conditions~\cite{resScalar}
\begin{equation}
  \label{OSren}
 \Pi_{\alpha\beta}^{\rm disp} (M^2_\alpha)\ =\
 \Pi_{\alpha\beta}^{\rm disp} (M^2_\beta)\ =\ 0\; ,\qquad
\lim_{s\to M^2_\alpha}\,
\frac{\, \Pi_{\alpha\alpha}^{\rm disp} (s) }{s -  M^2_\alpha}\ =\ 0\; .
\end{equation}
The  advantage of the  OS scheme  is that  around the  resonant region
$s\approx  M_1^2   \approx  M_2^2$,  the   OS-renormalized,  UV-finite
dispersive  self-energies  are  negligible,  and only  the  absorptive
self-energies $\Pi_{\alpha\beta}^{\rm abs}(s)$ become relevant. Hence,
one may obtain a Born-improved approximation for the propagator matrix
given by
\begin{eqnarray}
 {\Delta}^{-1} (s) &=& 
\left(
\begin{array}{cc}
 s\: -\: M_1^2\: +\: i\, \Pi_{11}^{\rm abs}(s) & i\,\Pi_{12}^{\rm abs}(s)\\[2mm]
 i\, \Pi_{21}^{\rm abs}(s) & s\: -\: M_2^2\: +\: i\, \Pi_{22}^{\rm abs}(s)
 \end{array} \right)\; .
\label{invProp}
\end{eqnarray}
Employing the standard Cutkosky cutting rules~\cite{CutRules} as shown
in Fig.~\ref{Fig:SiSj},  we can calculate  the absorptive part  of the
self-energies,  $\Pi_{\alpha\beta}^{\rm  abs}(s)$.   For  our  generic
theory   with  scalar  interactions   $\widetilde  S_\alpha$-$f_i$-$\tilde
f_{\tilde j \,}$, we obtain at the one-loop level,
\begin{eqnarray}
 \lefteqn{
\Pi_{\alpha\beta}^{\rm abs}(s)\ = \
\frac{1}{16 \pi s}\;\lambda^{1/2} (s,m^2_i,m^2_{\tilde j})
\; \theta(\sqrt s -m_i -m_{\tilde j}) } \nonumber\\[3mm] 
&&
\times\,  \Big[ \left(\CLs\CLb + \CRs\CRb\right) (s -m_{i}^2- m_{\tilde j}^2 )
         -2\,\left(\CRs\CLb + \CLs\CRb\right) m_{i} m_{\tilde j}
 \Big]\; ,\; 
\qquad 
\end{eqnarray}
where $\lambda (x,y,z) =  (x - y - z)^2 - 4yz$  and summation over all
intermediate OS states  $f_i \, \tilde f_{\tilde j  \,}$ is implicitly
assumed.      Observe     that     the    absorptive     self-energies
$\Pi_{\alpha\beta}^{\rm  abs}(s)$ are  related  to the  anti-Hermitian
part of the inverse propagator matrix~(\ref{invProp}), through
\begin{eqnarray}
\Pi_{\alpha\beta}^{\rm abs} (s) &=&
-\ \frac{i}{2} \, \Big[ \Delta^{-1}(s)\ -\ \big(\Delta^{-1}(s)\big)^\dagger 
\Big]_{\alpha\beta} \; . 
\end{eqnarray}
The latter is also related through the optical theorem to 
\begin{eqnarray}
  \label{OT}
\Big[ \Delta^{-1}(s)\ -\ \big(\Delta^{-1}(s)\big)^\dagger 
\Big]_{\alpha\beta}\ =\ 
i \, \sum_X \int {\rm dPS}_X \, V_X^\alpha  \, (V_X^\beta)^\dagger\; ,
\end{eqnarray}
where the  sum is over all  OS intermediate states  $X$.  In addition,
$V_X^{\alpha,\beta}$ are the vertex amplitudes for the decay processes
$\widetilde{S}^{\alpha,\beta}  \to   X$,  and  ${\rm   PS}_X$  is  the
respective  Lorentz-invariant phase  space.  Further  constraints from
unitarity and CPT invariance will be discussed in the next section.

%
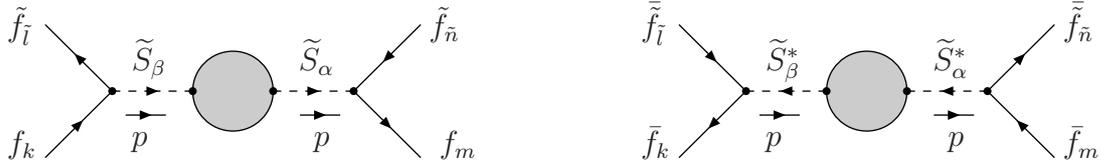
\begin{figure}[t]
\begin{minipage}{0.45\textwidth}
\begin{picture}(10,5)(-1,0)
	\put( -0.5,3.55){$\tilde f_{\tilde l}$}
	\put( -0.5,2.){$f_{k}$}
\ArrowLine(25,80)(0,105)
\ArrowLine(0,55)(25,80)
	\Vertex(25,80){1.5}
        \DashArrowLine(25,80)(55,80){3}
	\ArrowLine(30,71)(45,71)
        \put( 1.,2.1){ $p$}
         \GCirc(70,80){15}{0.80}
	\Vertex(55,80){1.5}
	\Vertex(85,80){1.5}
	\DashArrowLine(85,80)(115,80){3}
	\ArrowLine(95,71)(110,71)
        \put( 3.4,2.1){ $p$}
	\put( 1.0, 3.1){ $\widetilde S_\beta $}
	\put( 3.2, 3.1){ $\widetilde S_\alpha $}
       \Vertex(115,80){1.5}
\ArrowLine(140,105)(115,80)
\ArrowLine(115,80)(140,55)
	\put( 5.05,2.){   $f_{m}$}
	\put( 5.05,3.55){$\tilde f_{\tilde n}$}
\end{picture}
\end{minipage}
\begin{minipage}{0.45\textwidth}
\begin{picture}(10,5)(-2,0)
	\put( -0.5,3.55){$\bar{\tilde f_{\tilde l \,}}$}
	\put( -0.5,2.){$\bar f_k$}
\ArrowLine(0,105)(25,80)
\ArrowLine(25,80)(0,55)
	\Vertex(25,80){1.5}
        \DashArrowLine(55,80)(25,80){3}
	\ArrowLine(30,71)(45,71)
        \put( 1.,2.1){ $p$}
         \GCirc(70,80){15}{0.80}
	\Vertex(55,80){1.5}
	\Vertex(85,80){1.5}
	\DashArrowLine(115,80)(85,80){3}
	\ArrowLine(95,71)(110,71)
        \put( 3.4,2.1){ $p$}
	\put( 1.0, 3.1){ $\widetilde S_\beta^\ast $}
	\put( 3.2, 3.1){ $\widetilde S_\alpha^\ast $}
       \Vertex(115,80){1.5}
\ArrowLine(115,80)(140,105)
\ArrowLine(140,55)(115,80)
	\put( 5.05,3.55){$ \bar{\tilde f_{\tilde n}}$}
	\put( 5.05,2.){$\bar f_{m} $}
\end{picture}
\end{minipage}
\caption{\it Transition  amplitude $f_k \tilde f_{\tilde  l} \to f_{m}
  \tilde  f_{\tilde n}  $  via the  scalars $\widetilde  S_{\alpha,\beta}$
  (left), and its CP-conjugate amplitude (right). The grey circles
  indicate one-loop resummed contributions. }
\label{Fig:Strans}
\end{figure}

It is now  instructive to consider $2 \to  2$ scatterings, mediated by
the   coloured   scalar   particles  $\widetilde{S}^\alpha$   in   the
$s$-channel,  as shown  in  Fig.~\ref{Fig:Strans}.  Specifically,  the
resonant $s$-dependent part of the amplitude $A\to B$ is given by
\begin{eqnarray}
  \label{Ts}
 {\cal T}_s^{A\to B} (s) &=& 
  \sum_{\alpha,\beta}\ (V_B^\alpha)^\ast \: \Delta_{\alpha\beta}(s)\:
  V_A^\beta\ \equiv \ V_B^\dagger \: \Delta(s) \: V_A\; ,
\end{eqnarray}
where  we use  matrix notation  in the  space of  the  unstable scalar
particles  $\widetilde{S}^\alpha$ in  the last  step of  the equation.
Moreover, the  vertex functions, for the  in-state $A =  f_k \, \tilde
f_{\tilde l}  $ and the out-state $B  = f_m \, \tilde  f_{\tilde n} $,
have the analytic forms
\begin{eqnarray}
V_{A}^{\beta} \;\;\;\equiv \quad V_{k\tilde l}^{\beta} \quad &=& \;\;
       \bar v(p_{\tilde l}, \lambda_{\tilde l})\left(
                                  C^{\beta,R\ast}_{k\tilde l} P_L + 
                                  C^{\beta,L\ast}_{k\tilde l} P_R 
                                          \right)u(p_k, \lambda_k)\; ,\\[1mm]
(V_{B}^{\alpha})^\ast \;\equiv\; (V_{m\tilde n}^{\alpha})^\ast &=&
             \bar u(p_{m}, \lambda_m)\left(
                                  C^{\alpha,L}_{m\tilde n} \,P_L + 
                                  C^{\alpha,R}_{m\tilde n} \,P_R   
                                     \right) v(p_{\tilde n},
                                     \lambda_{\tilde n})\; ,
\end{eqnarray}
where $p_k$ and $\lambda_k$  symbolize the 4-momentum and the helicity
for  a  given fermion  $f_k$.   It~now  proves  useful to  define  the
Hermitian matrix
\begin{eqnarray}
  \label{Tab}
T_{A}^{\beta\gamma}(s)    &\equiv&
\sum_{\lambda_k,\lambda_{\tilde l}}
V_{A}^\beta \,(V_{A}^\gamma)^\ast \ =\
  \left(C^{\beta,L\ast}_{k\tilde l}C^{\gamma,L}_{k\tilde l}  +
        C^{\beta,R\ast}_{k\tilde l}C^{\gamma,R}_{k\tilde l} \right)  
        (s -m_{k}^2- m_{\tilde l}^2 )\nonumber\\
&& \hspace{3.2cm} -\ 2\left(C^{\beta,R\ast}_{k\tilde l}C^{\gamma,L}_{k\tilde l}  +
       C^{\beta,L\ast}_{k\tilde l}C^{\gamma,R}_{k\tilde l} 
 \right)m_{k}m_{\tilde l}\; , \qquad
\end{eqnarray}
for   the  production  vertex,   including  an   analogous  definition
$T_{B}^{\beta\gamma}$  for  the  decay  vertex.   In  terms  of  these
matrices, the  squared amplitude for the  $s$-channel resonant process
$A \to \widetilde{S}_\alpha \to B$ takes on the simple form:
\begin{eqnarray}
  \label{traceAB}
|{\cal T}_s^{ A\to B}(s)|^2 &=&
T^{\delta\alpha}_{B}(s) \: \Delta_{\alpha\beta}(s)\:
T^{\beta\gamma}_{A}(s) \: \Delta_{\gamma\delta}^\dagger(s)
\ = \    {\rm Tr}\Big[  T_{B}(s) \, \Delta (s)\, T_{A}(s) \,
  \Delta^\dagger(s)\Big]\; .\quad 
\end{eqnarray}
Notice that the squared amplitude can be compactly written in the form
of a trace over the scalar-particle space $\widetilde{S}_\alpha$.

For the  CP-conjugate process $\bar A  \to \widetilde{S}^*_\alpha \to
\bar B$, the transition amplitude may be written down as follows:
\begin{eqnarray}
{\cal T}_s^{\bar A \to \bar B}\  =  \
    (\overline V_{B}^{\alpha})^\ast \: \Delta^{\sf T}_{\alpha\beta}(s)\: 
     \overline V_{A}^{\beta}\ = \ \overline V_B^\dagger \: \Delta^{\sf T} (s) \:
    \overline V_A\; ,
\end{eqnarray}
where the CP-conjugate in- and out-state vertex functions are given by
\begin{eqnarray}
 (\overline V_{B}^{\alpha})^\ast   &=&
       \bar v(\bar p_{{\tilde n}}, \lambda_{{\tilde n}})\left(
                                  C^{\alpha,R\ast}_{m\tilde n} P_L + 
                                  C^{\alpha,L\ast}_{m\tilde n} P_R 
                                \right)u(\bar p_{m}, \lambda_{m})\ 
= \ V_{B}^{\alpha}(\bar p) \; ,\\[1mm]
\overline V_{A}^{\beta} &=&
\bar u(\bar p_{ k}, \lambda_{k})\left(
                                  C^{\beta,L}_{k\tilde l} \,P_L + 
                                  C^{\beta,R}_{k\tilde l} \,P_R   
        \right) v(\bar p_{{\tilde l}}, \lambda_{{\tilde l}} )
\ =  \ \big[ V_{A}^{\beta}(\bar p)\big]^\ast\; ,
\end{eqnarray}
with $\bar p = (E, -{\mathbf  p})$. The RHSs of the last two equations
give   the   relations  of   the   CP-conjugate  vertices   $\overline
V_{A,B}^{\alpha}$  to $V_{A,B}^{\alpha}$.  Taking  these relationships
into account, we can express the squared amplitude of the CP-conjugate
process entirely in the $\widetilde{S}_\alpha$-space as follows:
\begin{eqnarray}
  \label{traceABCP}
|{\cal T}_s^{ \bar A \to \bar B}(s)|^2 &=&
\sum_{\lambda_{k},\lambda_{m}, 
      \lambda_{{\tilde l}},\lambda_{{\tilde n}}}
 \overline V_{B}^\dagger \, \Delta^{\sf T} \,  \overline V_{A}  \
 \overline V_{A}^\dagger \, (\Delta^{\sf T})^\dagger \, \overline V_{B} \\[2mm]
 &=&
{\rm Tr}\Big[ T_{B}^{\sf T}(s)\: \Delta^{\sf T}(s)\: T_{A}^{\sf T}(s)\:
  \big(\Delta^{\sf T}(s)\big)^\dagger \Big]\  =\ 
 {\rm Tr}\Big[  T_A(s) \: \Delta(s)\: T_B(s) \: 
                                  \Delta^\dagger (s) \Big]\; .\nonumber
\end{eqnarray}
In fact, the last two  trace expressions correspond effectively to the
discrete transformations CP and T-reversal, respectively.  Explicitly,
under  CP-conjugation, the  resummed $\widetilde{S}_\alpha$-propagator
matrix and the production and decay matrices transform as
\begin{equation}
\mbox{CP:}\quad \Delta (s) \quad \to \quad \bar{\Delta}(s)\ =\
\Delta^{\sf T}(s)\;;\qquad  T_{A\,(B)}(s) \quad \to \quad
\overline{T}_{A\, (B)}(s)\ =\ T^{\sf T}_{A\, (B)}(s)\; .
\end{equation}
Under T-reversal, we have
\begin{equation}
\mbox{T:}\quad \Delta (s) \quad \to \quad \Delta^{\rm t}(s)\ =\
\Delta (s)\;;\qquad  T_{A,B}(s) \quad \to \quad
T^{\rm t}_{A\, (B)}(s)\ =\ T_{B\, (A)}(s)\; .
\end{equation}
Notice      that      the      squared      amplitudes~(\ref{traceAB})
and~(\ref{traceABCP})    are    manifestly    invariant   under    CPT
transformations.

\setcounter{equation}{0}
\section{Unitarity and CPT Constraints on CP Observables}
\label{CPTobservables}

The unitarity and CPT invariance of the S-matrix introduce non-trivial
constraints   on   CP   asymmetries,  see~e.g.~\cite{resScalar}.    In
particular, the  equality of the $s$-channel forward  amplitude $A \to
A$ with its CPT-conjugate one $\bar{A} \to \bar{A}$,
\begin{eqnarray}
  \label{eq:CPT}
{\cal T}_s^{A\to A}(s) &=&  {\rm Tr}\Big[ T_A (s) \, \Delta (s) \Big]
\ =\ {\cal T}_{s}^{\bar A\to \bar A}(s)\; ,
\end{eqnarray}
gives  rise to  relations  among the  different  resonant channels  in
processes, such as $A \to X$ and its CP-conjugate process $\bar{A} \to
\bar{X}$, where $X$ ($\bar{X}$) is an accessible final state.  To make
such relations  explicit, let us  first consider the reactions  of $A$
into  all possible  $X$  states.  The sum  of  the squared  amplitudes
integrated over the phase space of the $X$ states may be calculated as
\begin{eqnarray}
  \label{eq:OT}
\sum_X \int {\rm dPS}_X\, | {\cal T}_{s}^{A\to X}|^2 &=&
\sum_X {\rm PS}_X\; {\rm Tr}\Big[  T_{A} \, \Delta \,  T_{X}
\Delta^\dagger \Big]\ =\
-i \, {\rm Tr}\Big[  T_{A} \, \Delta  \, 
        \Big( \Delta^{-1} - (\Delta^{-1})^\dagger\Big)
       \,  \Delta^\dagger \Big]\nonumber\\[2mm]
&=&
i \, {\rm Tr}\Big[\,T_{A} \, \Big(\Delta\ -\  
                             \Delta^\dagger\Big)\, \Big]
\ =\
-\, 2\, {\rm Im} \Big( {\cal T}_{s}^{A\to A} \Big)\; ,
\end{eqnarray}
where  we  used  the  relation~(\ref{OT})  derived  from  the  optical
theorem.  As  a consequence of the  CPT constraint~(\ref{eq:CPT}), the
total unpolarised cross section of $A$ to all possible $X$ is equal to
the corresponding unpolarised cross  section of the C-conjugate states
$\bar{A}$ to all possible C-conjugate states $\bar{X}$.
 
Assuming  that  the  dominant  source   of  CP  violation  is  due  to
scalar-particle mixing,  we define CP-violating  observables pertinent
to  a resonant  reaction $A  \to \widetilde{S}_\alpha  \to B$  and its
CP-conjugate one $\bar{A} \to \widetilde{S}^*_\alpha \to \bar{B}$:
\begin{eqnarray}
  \label{CPobs}
\Delta^{{\rm CP}}_{A\to B}(s) &=&
\int \Big( |{\cal T}_s^{A\to B} (s)|^2 - 
                         |{\cal T}_s^{\bar A\to \bar B} (s)|^2   \Big) 
        \,{\rm dPS}_B\; .
\end{eqnarray}
Since  we sum  over all  particle  helicities and  integrate over  the
P-invariant   phase  space,   the   CP-odd  observable   $\Delta^{{\rm
    CP}}_{A\to  B}$   is  also   odd  under  C,   but  even   under  P
transformations.

We will now show that thanks to unitarity and CPT invariance, at least
three linearly independent decay  channels of $\widetilde S_{\alpha }$
are  required in  $2\to 2$  scatterings, in  order to  obtain non-zero
C/CP-odd  observables, such as  $\Delta^{{\rm CP}}_{A\to  B}(s)$ given
in~(\ref{CPobs}). Our  proof for  a two-particle scalar  mixing system
will then be generalized to  mixing systems with $n$ scalar particles.
In  Section~\ref{sec:EPR}, we  extend this  theorem to  the correlated
production  and decay  of two  unstable-particle systems  in  $2\to 4$
scatterings, where  we show that  only two independent  decay channels
are sufficient to obtain a non-zero C and CP asymmetry.

\subsection{CP Conditions in Two-Particle-Mixing Transitions}
\label{CPcond2mix}

We  will now  show that  the C-  and CP-odd  observables $\Delta^{{\rm
    CP}}_{A\to  B}(s)$ defined  in~(\ref{CPobs}) vanish,  if  only two
independent decay  channels are open in $2\to  2$ resonant scatterings
$A \to \widetilde{S}_\alpha  \to X$.  To this end,  let us assume that
$X  = A',\,  B$, where  $A'$  is the  same two-particle  state as  the
initial  state $A$ but  has different  momentum configurations.   As a
consequence   of   the    CPT-invariance   and   unitarity   relations
in~(\ref{eq:CPT}) and~(\ref{eq:OT}), we have the equality
\begin{equation}
\int {\rm dPS}_{A^\prime}\,|{\cal T}_s^{A \to A^\prime}|^2\; +\: 
\int {\rm dPS}_B\,|{\cal T}_s^{ A \to B}|^2 
\ =\
\int {\rm dPS}_{ A^\prime}\,|{\cal T}_s^{ \bar A \to \bar A^\prime}|^2\; 
+\: 
\int {\rm dPS}_{B}\,|{\cal T}_s^{\bar A \to \bar B}|^2 \; .\quad
\end{equation}
In terms of CP observables, the above equality can be rewritten as
\begin{eqnarray}
\Delta^{{\rm CP}}_{A\to B}(s)
&=& 
\phantom{-}\int {\rm dPS}_{B}\left(
    |{\cal T}_s^{     A \to      B}|^2
   -|{\cal T}_s^{\bar A \to \bar B}|^2
                  \right)\ =\ 
-\int {\rm dPS}_{A^\prime}\left(
    |{\cal T}_s^{     A \to      A^\prime}|^2
   -|{\cal T}_s^{\bar A \to \bar A^\prime}|^2
                  \right)\nonumber\\
&=& -\, \Delta^{{\rm CP}}_{A \to A }(s)\; .
\end{eqnarray}
Since the decay channel  of $\widetilde{S}_\alpha$ into states $A'$ is
not  kinematically  independent  to  the  production  channel  $A  \to
\widetilde{S}_\alpha$  (they  just differ  by  an overall  phase-space
factor),  we  have  that $\Delta^{{\rm  CP}}_{A  \to  A}  = 0$,  as  a
consequence of CPT invariance, implying also that $\Delta^{{\rm CP}}_{A
\to B}(s) = 0$. Hence,  we can conclude  that more  than two  linearly
independent decay channels would be needed to obtain non-zero C and CP
asymmetries in $2\to 2$ resonant scatterings.

The above conclusion  can also be obtained by  an explicit calculation
of  the squared  amplitude  $A\to \widetilde{S}_\alpha  \to  B$, in  a
two-particle mixing  system; the generalization  to a system  with $n$
particles  follows  in  the  next subsection.   Our  calculation  gets
facilitated by  a spinorial  trace technique, where  the $2  \times 2$
production  and decay  matrices $T_{A,B}$  and the  inverse propagator
matrix  $\Delta^{-1}$ are  expanded in  terms of  the  two-by-two unit
matrix  $\One_2 \equiv  \sigma^0$  and  the  three Pauli  matrices
$\sigma^{1,2,3}$. Technical  details of the trace  technique are given
in  Appendix~\ref{sec:Matrixformalism}.  In   detail,  we  obtain  for
the squared amplitude
\begin{eqnarray}
   \label{evaluatetrace}
|{\cal T}_s^{ A \to  B}|^2 &=&
{\rm Tr}\left[ T_A \,\Delta \, T_B \, \Delta^\dagger \right]
\ =\
\frac{ T_{A}^\mu   \,\Delta^{-1,\nu} \, 
       T_{B}^\rho \, (\Delta^{-1,\lambda})^\ast }
      {|\det(\Delta^{-1})|^2 }\
{\rm Tr}\left[ \sigma_{\mu}\bar\sigma_{\nu}
\sigma_{\rho}\bar\sigma_{\lambda}\right]\; ,
\end{eqnarray}
where     the     expansion      coefficients     are     given     in
Appendix~\ref{sec:exPropagatormatrix}.    Since    the    CP-conjugate
amplitude  is obtained  by  the interchange  of  the Hermitian  vertex
matrices $T_A  \leftrightarrow T_B$, the  only contribution to  the CP
observable  $\Delta^{{\rm  CP}}_{A\to   B}(s)$  comes  from  the  $\mu
\leftrightarrow   \rho$   antisymmetric   parts   of   the   amplitude
squared.  These are  the terms  proportional to  the  the Levi--Civita
tensor   $\varepsilon_{\mu\nu\rho\lambda}$  contained  in   the  trace
expression   over   the  generalized   Pauli   spinors   on  the   RHS
of~(\ref{evaluatetrace})  [cf.~(\ref{4sigmatrace})].  Specifically, up
to  overall   phase-space  and  other  kinematical   factors,  the  CP
observable $\Delta^{{\rm CP}}_{A\to B}(s)$ is
\begin{equation}
  \label{DCPAtoB}
  \Delta^{{\rm CP}}_{A\to B}(s) \ \propto\
\varepsilon_{\mu\nu\rho\lambda}\ T_A^\mu  \, T_B^\nu \, 
{\rm Im} (\Delta^{-1,\rho})\:  {\rm Re} (\Delta^{-1,\lambda})\; .
\end{equation}
The  4-vector  ${\rm  Im} (\Delta^{-1,\rho})$  derives  from
absorptive parts of the self-energy, 
\begin{equation}
{\rm Im}( \Delta^{-1,\rho} ) \ =\ 
\frac{1}{2}\ {\rm Tr} \left[\bar{\sigma}^\rho\, \Pi^{\rm abs}\right]\ =\ 
\frac{1}{2}\; \Big(\,{\rm PS}_A\, T^\rho_A\: +\: {\rm PS}_B\,
T^\rho_B\, \Big)\; , 
\end{equation}
where ${\rm PS}_{A,B}$ denote  phase-space factors associated with the
$A$ and $B$ states.  Evidently, three linearly independent channels or
4-vectors   in  the   generalized  Pauli-spinor   space   are  needed,
e.g.~$T^\rho_{A,B,C}$, in  order to obtain a  non-zero CP asymmetry
$\Delta^{{\rm CP}}_{A\to B}$ in~(\ref{DCPAtoB}).

\subsection{Beyond the Two-Particle Mixing}\label{Beyond2mix}

The above result obtained for  a two-particle-mixing system can now be
generalized   to   mixing   systems   with  $n$   unstable   particles
$\widetilde{S}_\alpha$, with $\alpha = 1,2,\dots, n$.

Our  starting point  is  the $n\times  n$  propagator matrix  $\Delta$
given by
\begin{equation}
\Delta (s)\ =\ \left[{\rm D}^{-1}(s)\: +\: i \,\Pi(s)\right]^{-1}\; .
\end{equation}
The $n\times n$ propagator matrix $\Delta (s)$ consists of a Hermitian
matrix  ${\rm  D}_{\alpha\beta}^{-1}(s)  = \delta_{\alpha\beta}  (s  -
M_\alpha^2)  +  \Pi_{\alpha\beta}^{\rm  disp}(s)$, which  may  include
dispersive   contributions,  and   a  general   anti-Hermitian  matrix
$i\,\Pi_{\alpha\beta}(s)  = i\,  \Pi_{\alpha\beta}^{\rm  abs}(s)$ that
describes   the    absorptive   effects.    Dropping    the   explicit
$s$-dependence, it can be rewritten in the more convenient form:
\begin{equation}
\Delta\ =\ ( {\rm D} - i \,{\rm D}\, \Pi \, {\rm D})
     \left[    \One + (\Pi \, {\rm D})^2 \right]^{-1}\ =\ 
\left[    \One + ( {\rm D} \, \Pi )^2 \right]^{-1}
     ( {\rm D} - i \,{\rm D}\, \Pi \, {\rm D})\; ,
\end{equation}
where $\One$ is the $n$-dimensional unit matrix.

The  cross  sections  for  the  $2  \to 2$  resonant  process  $A  \to
\widetilde{S}_\alpha  \to B$  and  its CP-conjugate  one $\bar{A}  \to
\widetilde{S}^*_\alpha \to \bar{B}$ may straightforwardly be evaluated
by the trace expressions
\begin{equation}
\sigma\ = \  {\rm Tr}\left[ T_B \, \Delta  \ T_A \, \Delta^\dagger  \right]\;,
\qquad
\sigma^{\rm CP}\ =\  {\rm Tr}\left[ 
T_A \, \Delta  \ T_B \, \Delta^\dagger  \right]\; ,
\end{equation}
where   phase-space   integration  is   implied.    From  these,   the
CP-violating difference of the two cross sections, $\Delta \sigma^{\rm
  CP}  \equiv \sigma  - \sigma^{\rm  CP}$,  can be  defined, which is
calculated as
\begin{eqnarray}
  \label{yzz}
\Delta \sigma^{\rm CP} &=& 
\frac{1}{2}\;
{\rm Tr}\Big[ T_B \, (\Delta - \Delta^\dagger) \,
             T_A \, (\Delta + \Delta^\dagger)\  
            -\ T_B \, (\Delta + \Delta^\dagger) \,
             T_A \, (\Delta - \Delta^\dagger)  \Big]\nonumber\\[3mm]
&=& -2i \, {\rm Tr} \Big\{ T_B \, {\rm D} \, \Pi  \, 
                [\One + ( {\rm D} \, \Pi)^2  ]^{-1} \, 
                 {\rm D}  \, T_A \,
                [\One + ( {\rm D}\, \Pi )^2  ]^{-1} \, {\rm D}\nonumber\\
& & \hspace{1.5cm}
                -\  T_B \, 
                [\One + ( {\rm D} \, \Pi)^2  ]^{-1} \,
                 {\rm D} \, T_A \, {\rm D} \, \Pi  \,
                [\One + ( {\rm D} \, \Pi)^2  ]^{-1} \, {\rm D}
            \Big\}\; .
\end{eqnarray}
From  the   last  expression  in~(\ref{yzz}),  we   observe  that  the
CP-violating difference $\Delta \sigma^{\rm  CP}$ can be cast into the
compact form:
\begin{equation}
  \label{CPcommutator}
\Delta\sigma^{\rm CP}\ =\ 
-2i\, {\rm Tr} \Big\{  {\rm D} \, \Pi  \, 
        \Big[
   \left( \One + ( {\rm D} \, \Pi)^2  \right)^{-1} \, {\rm D} \, T_A\:
   ,  \:
   \left( \One + ( {\rm D} \, \Pi)^2  \right)^{-1} \, {\rm D} \, T_B 
        \Big]
         \Big\}\; ,
\end{equation}
where $[X\,,\,Y] \equiv  XY-YX$ is the commutator for  two $n\times n$
matrices  $X$  and $Y$.   Clearly,  if  $T_B =  c  T_A$,  with $c  \in
\mathbb{C}   $,  the   commutator   in~(\ref{CPcommutator})  vanishes.
Further, from the cyclicity of the trace,
\begin{eqnarray}
{\rm Tr} \left\{  A\, [B\, ,\,C] \right\}\ =\
{\rm Tr} \left\{  C\, [A\, ,\,B] \right\}\ =\
{\rm Tr} \left\{  B\, [C\, ,\,A] \right\},
\label{traceProp}
\end{eqnarray}
we    have     that    the    CP-violating     difference    vanishes,
i.e.~$\Delta\sigma^{\rm CP} = 0$, if
\begin{eqnarray}
\Big[
   \left( \One + ( {\rm D} \, \Pi)^2  \right)^{-1} \, {\rm D} \, T_A\: ,\:
      {\rm D} \, \Pi  \,   
  \Big] &=& 0\; , 
\end{eqnarray}
or if
\begin{eqnarray}
\Big[
   \left( \One + ( {\rm D} \, \Pi)^2  \right)^{-1} \, {\rm D} \, T_B\: ,\:
      {\rm D} \, \Pi  \, \Big] &=& 0\; .
\end{eqnarray}
Since the  absorptive contributions can in general  be parametrized by
the  different open  decay channels,  $\Pi =  \alpha T_A  +  \beta T_B
+\gamma T_C$,  then $\Delta\sigma^{\rm CP}$  will vanish if $T_C  = 0$
(only 2  open decay  channels), or  if $T_C =  c_1 T_A  + c_2  T_B$ is
linearly dependent.  Consequently, at least three linearly independent
decay channels are needed to construct non-vanishing CP asymmetries in
general $2  \to 2$ scattering  processes. As we  will see in  the next
section,  the validity  of this  theorem may  be extended  to unstable
states which are  pair produced and decay in correlation  in $2 \to 4$
scattering reactions, where however  only two independent channels are
needed.

We  conclude  this section  by  remarking that  in  the  absence of  a
particle  mixing,  the  decay   matrices  $T_{A,B,C}$  and  $\Pi$  are
diagonal, and the commutator in~(\ref{CPcommutator}) vanishes, leading
to $\Delta\sigma^{\rm CP}=0$.  This corresponds to the case where each
unstable particle couples to a disjoint set of states.  This situation
can happen naturally,  only if the unstable particles  carry their own
conserved charges.  In minimal  supersymmetric theories, the mixing of
light squarks or sleptons is  suppressed by their masses, such that CP
asymmetries are  sizable, only  if particular resonant  conditions are
met~\cite{Marek,resScalar,nima}.  As we will  see in the next section,
we find sizable  CP asymmetries in the strongly  mixed stop or sbottom
sectors for a wide range of MSSM parameters.

\setcounter{equation}{0}
\section{EPR Correlated Production and Decay of Scalars}
\label{sec:EPR}

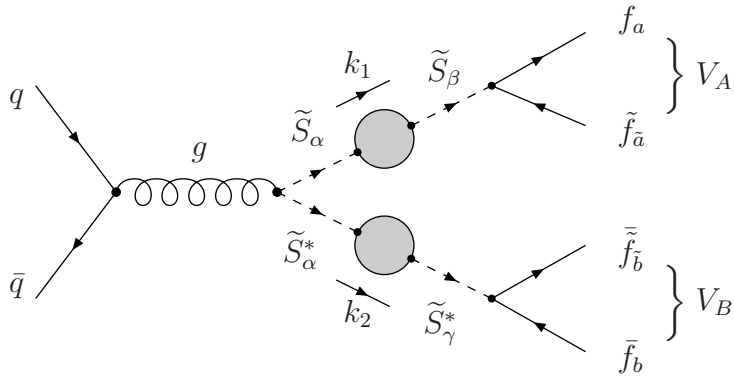
\begin{figure}[t]
\scalebox{1}{
\begin{picture}(10,5)(-3.4,0)
	\ArrowLine(-40,120)(-10,80)
	\ArrowLine(-10, 80)(-40,40)
	\Vertex(-10,80){1.8}
	\Gluon(-10,80)(50,80){5}{5}
	\Vertex(50,80){1.8}
	\put( -1.9,4.0){ $q$}
	\put( -1.9,1.5){ $\bar q $}
	\put( 0.5,3.3){ $g $}
         \GCirc(90,100){11}{0.80}
	\Vertex( 80, 95){1.5}
	\Vertex(100,105){1.5}
        \DashArrowLine(50,80)(80,95){3}
        \DashArrowLine(100,105)(130,120){3}
	\Vertex(130,120){1.5}
	\ArrowLine(130,120)(165,140)
	\ArrowLine(165,105)(130,120)
	\put( 1.8,3.5){ $\widetilde S_\alpha$}
	\put( 6.1,5.0){ $f_a$}
	\put( 6.1,3.5){ $\tilde f_{\tilde a} $}
	\put( 6.7,4.25){ $ \Bigg\} \; V_A$}
	\ArrowLine(72,112)(92,122)
        \put( 2.5,4.4){ $k_1$}
	\put( 3.6,4.3){ $\widetilde S_\beta$}
         \GCirc(90,60){11}{0.80}
	\Vertex( 80, 65){1.5}
	\Vertex(100, 55){1.5}
        \DashArrowLine(50,80)(80,65){3}
        \DashArrowLine(100,55)(130,40){3}
 	\Vertex(130,40){1.5}
	\ArrowLine(130,40)(165,60)
	\ArrowLine(165,20)(130,40)
	\put( 1.7,1.9){ $\widetilde S_\alpha^\ast $}
	\put( 6.1,0.5){ $\bar f_b$}
	\put( 6.1,1.9){ $\bar{\tilde f_{\tilde b}}$}
	\put( 6.7,1.25){ $ \Bigg\} \; V_B$}
	\ArrowLine(72,47)(92,37)
        \put( 2.5,1.1){ $k_2$}
	\put( 3.55,0.95){ $\widetilde S_\gamma^\ast$}

\end{picture}}
\caption{\it Correlated production and decay of $\widetilde S_\alpha$
  particles in $2\to 4$ scatterings at high-energy hadron colliders.}
\label{Fig:decayStop}
\end{figure}

In this  section we study  the EPR-correlated production and  decay of
unstable  scalar particles  at high-energy  colliders.  Specifically, we
consider the resonant part  of $2\to4$ processes, mediated by coloured
or  charged scalar  particles~$\widetilde S_\alpha$,  as  displayed in
Fig.~\ref{Fig:decayStop}.  

The transition  amplitude for the $2\to 4$  partonic process $q\bar{q}
\to  \widetilde{S}_\alpha \widetilde{S}^*_\alpha  \to (f\,\tilde{f})\:
(\bar{f}'\,\bar{\tilde{f}'})$ has the analytical form
\begin{eqnarray}
{\mathcal T}_s^{2\to 4}(k_1,k_2) &=& 
 J_g(k_1,k_2) \, V_A^\dagger(k_1)\,
 \Delta(s_1)\,\Delta(s_2)\,V_B(k_2)\; ,
\end{eqnarray}
where  the   initial  gluon  current  factorizes   in  the  amplitude,
$J_g(k_1,k_2) = J^\mu_g\,(k_1 -k_2)_\mu$~\footnote{Here we discuss the
  LHC suppressed, but more intuitive one-gluon exchange process. Other
  processes,   such   as   $gg\to   \widetilde   S_\alpha   \widetilde
  S_\alpha^\ast  \to  (f\,\tilde{f})\:  (\bar{f}'\,\bar{\tilde{f}'})$,
  exhibit similar analytical features  and can be treated in analogous
  fashion.}. Absorbing all kinematical factors into an overall
normalization constant ${\cal N}$,  the cross section
for the $2\to 4$ scattering is given by
\begin{equation}
\sigma^{2\to4}(s)\ =\ {\cal N}\: 
 {\rm Tr}\Big[\, T_A(s_1)\,\Delta(s_1)\,\Delta(s_2)\,
               T_B(s_2)\,\Delta^\dagger(s_2)\,\Delta^\dagger(s_1)  \,
          \Big]\; ,
\label{2to4scatamplitude}
\end{equation}
where  $s  = (k_1+k_2)^2\gg  M_\alpha^2$,  $s_{1,2}  = k^2_{1,2}$  and
phase-space integration is understood.

Since we are interested in the  dominant part of the cross section, we
expand $\sigma^{2\to4}(s)$ about the  resonant region $\sqrt{\bar s} =
(M_1 + M_2)/2$.   Hence, we obtain for the  resonant~(res) part of the
cross section
\begin{equation}
  \label{sigmares}
\sigma^{2\to4}_{\rm res}(s) \ =\ {\cal N}\:
{\rm Tr}\Big[\,T_A(\bar s)\,\left(\Delta(\bar s)\right)^2\, 
              T_B(\bar s)\,\left(\Delta^\dagger(\bar s)\right)^2\Big]\; ,
\label{eq:respart}
\end{equation}
where  we  have  ignored  sub-dominant  terms  proportional  to  $(s_1
-s_2)/\bar{s}$.  

To   linearize  the  expression~(\ref{sigmares})   in  terms   of  the
propagator  matrix  $\Delta  (\bar{s})$  and its  Hermitian  conjugate
$\Delta^\dagger(\bar{s})$, we define
\begin{equation}
  \label{Dxs}
\Delta(x;  s)\ \equiv\
\left[ x\One - 
        M_\alpha^2 +
           i \,\Pi^{\rm abs}_{\alpha\beta}(s) \right]^{-1}\; ,
\end{equation}
with $\Delta(s ; s) = \Delta(s)$, such that
\begin{equation}
  \label{D2xs}
\left(\Delta(s)\right)^2 \ =\ \left(\Delta(s; s)\right)^2\ =\ 
-\ \frac{\partial}{\partial x}\: \Delta(x; s)\Big|_{x=s}\; .
\end{equation}
With the help of this last relation, the resonant part of the $2\to 4$
cross section can now be recast into the form:
\begin{equation}
\sigma^{2\to4}_{\rm res}\ =\  {\cal N}\; 
\frac{\partial^2}{\partial x\: \partial y} \, 
{\rm Tr}\Big[\,T_A(\bar s)\,\Delta(x;\bar s)\, 
              T_B(\bar s)\,\Delta^{\dagger}(y;\bar s)\Big]
\Big|_{x=y=\bar s}\; .
\end{equation}
Note that  this form  is very  similar to the  squared amplitude  of a
$2\to 2$ resonant process [cf.~(\ref{evaluatetrace})].  
For the  specific two particle mixing  case, the resonant  part of the
$2\to 4$ cross section is given by
\begin{equation}
\sigma^{2\to4}_{\rm res}\ =\ {\cal N}\:
\frac{\partial^2}{\partial x\: \partial y} \ 
\frac{  \, \Delta^{-1,\nu}(x;\bar s)\, 
        \, \left(\Delta^{-1,\lambda}(y;\bar s) \right)^\ast}
     { \det [\Delta^{-1}(x;\bar s)]\: \det [\Delta^{-1}(y;\bar s)]^\ast}\,
 \Bigg|_{x=y=\bar s}
\! \! \!\! \!
\times \,
T_A^\mu  \, T_B^\rho \;
{\rm Tr} \left[ \sigma_\mu \bar \sigma_\nu \sigma_\rho 
\bar\sigma_\lambda\right]\; .\quad
\end{equation}
From this  last expression,  we may calculate  the CP-odd part  of the
 $2\to 4$ cross section, which may conveniently be written down as
\begin{equation}
(\sigma^{2\to4}_{\rm res})^{\sla{\rm{ CP}}}\ =\ 4\,{\cal N}\,
\varepsilon_{\mu\nu\rho\lambda}\,
T_A^\mu \, T_B^\nu\
{\rm Im}\left(\frac{\partial}{\partial x}\, 
\frac{ \Delta^{-1,\rho} (x;\bar s)}{ 
   \det [\Delta^{-1}(x;\bar s)]}\right)\: 
{\rm Re}\left(\frac{\partial}{\partial x}\, 
\frac{\Delta^{-1,\lambda} (x;\bar s)}{ 
   \det [\Delta^{-1}(x;\bar s)]}\right)\, ,
\end{equation}
where $x$ is set to  $\bar{s}$ after differentiation. Unlike in the $2
\to  2$ scatterings [cf.~(\ref{DCPAtoB})],  we may  convince ourselves
that  for $2\to 4$  scatterings, only  two linearly  independent decay
channels would be sufficient, i.e.
\begin{equation}
  \label{TAB2channels}
{\rm Im} \left(\Delta^{-1,\rho} (\bar s)\right)\ =\ 
\alpha\, T_A^\rho\ +\ \beta\, T_B^\rho \; ,
\end{equation}
in  order  to obtain  a  non-zero  CP-odd  contribution to  the  cross
section. Specifically, it is not difficult to see that the CP-odd
part $(\sigma^{2\to4}_{\rm res})^{\sla{\rm{ CP}}}$ contains terms
proportional to 
\begin{equation}
\varepsilon_{\mu\nu\rho\lambda}\,
T_A^\mu \, T_B^\nu\, \left. \frac{\partial\, {\rm Re}\,\Delta^{-1,\rho}
  (x;\bar s)}{\partial x}\, \right|_{x=\bar{s}}\;
{\rm Re}\, \Delta^{-1,\lambda} (\bar s)\
{\rm Im}\, \Big({\rm det}\,[\Delta (\bar s)]\Big)\; ,
\end{equation}
which   are  clearly  non-zero,   if  only   two  channels   as  given
in~(\ref{TAB2channels}) are assumed.

In our  discussion, we  have ignored other  one-loop effects,  such as
gluon-scalar-scalar   vertex  corrections.    These   corrections  are
sub-dominant,    but     become    relevant    in     restoring    the
gauge-fixing-parameter independence  of the gluon  propagator by means
of    Ward--Takahashi    identities     (for    a    recent    review,
see~\cite{reviewPT}).  In  particular, in  the Feynman gauge,  one can
show that  all these effects are  non-resonant and can  thus be safely
neglected.  A~detailed  study of these higher-order  effects is beyond
the scope of the present paper.

\subsection{Application to Stop CP Violation at the LHC}\label{sec:stopCP}

As an application of our  results, we consider the pair production and
decay     of     supersymmetric     scalar     top     (or     bottom)
quarks~\cite{mssm}. These coloured particles can be copiously produced
via  partonic  QCD interactions  at  hadron  colliders. Their  sizable
mixing can  lead to  large asymmetries, due  to non-vanishing  SUSY CP
phases in the heavy squark  sector $\tilde q$, $q=t,b$. Their dominant
decay   modes  are 
\begin{eqnarray}
\tilde q_{1,2}  &\to&  q + \tilde g\;,       \label{decaystop1}\\
\tilde q_{1,2}  &\to&  q + \tilde\chi_{1,2,3,4}^0\;, \label{decaystop2}\\
\tilde q_{1,2}  &\to&  q^\prime + \tilde\chi_{1,2}^+\;. \label{decaystop3}
\end{eqnarray}
Instead, the decays of $\tilde q_{1,2}$ into neutral bosons are closed
for sufficiently small $\tilde q_1$--$\tilde q_2$ mass splittings, i.e.
\begin{eqnarray}
\tilde q_2  &\slashed\to&  \tilde q_1 + H_{1,2,3}^0\;, \\
\tilde q_2  &\slashed\to&  \tilde q_1 + Z^0\; .
\end{eqnarray}
Likewise, their decays into charged bosons are closed for sufficiently
small stop and sbottom mass differences, i.e.
\begin{eqnarray}
\tilde q_{1,2}  &\slashed\to&  \tilde q^\prime_{1,2} + H^+\; , \\
\tilde q_{1,2}  &\slashed\to&  \tilde{q^\prime}_{1,2} + W^+\; .
\end{eqnarray}  
Stop  (and   sbottom)  decays  at   the  one-loop  level   within  the
CP-violating  MSSM have  been  calculated in~\cite{Heinemeyer:2010mm},
and in the CP-conserving MSSM in~\cite{Hlucha:2011yk}.

\begin{figure}[t]
\centering
\begin{picture}(16,15.5)
\put(-2.15,-5.9){\includegraphics{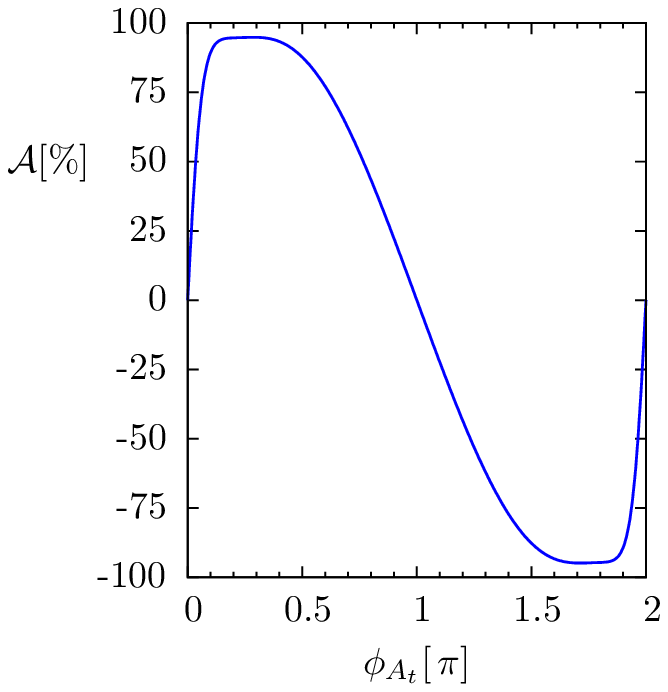}}
\put(6,-5.9){\includegraphics{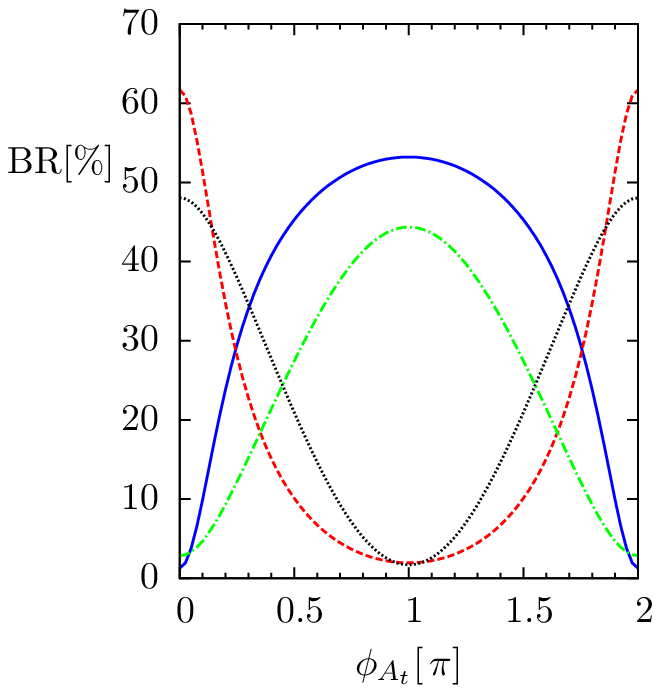}}
\put(1,8.15){(a)}
\put(9,8.15){(b)}
\put(-2.15,-13.9){\includegraphics{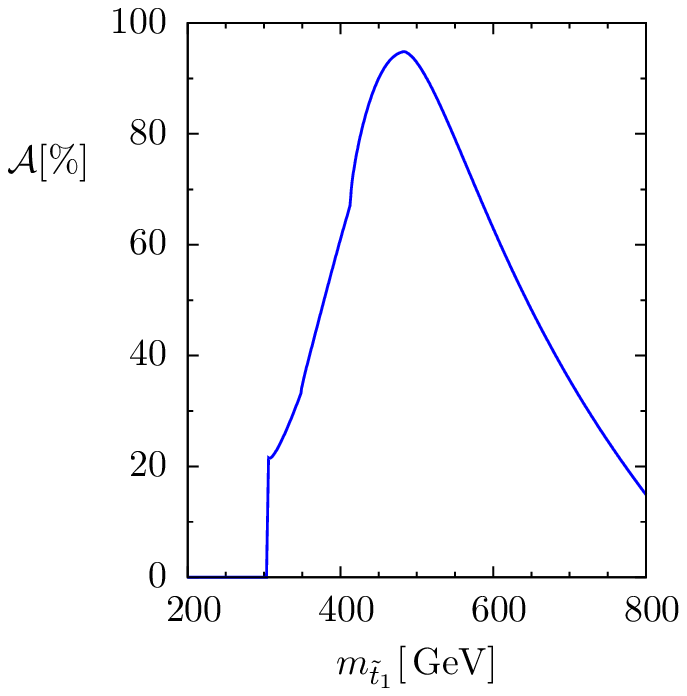}}
\put(6,-13.9){\includegraphics{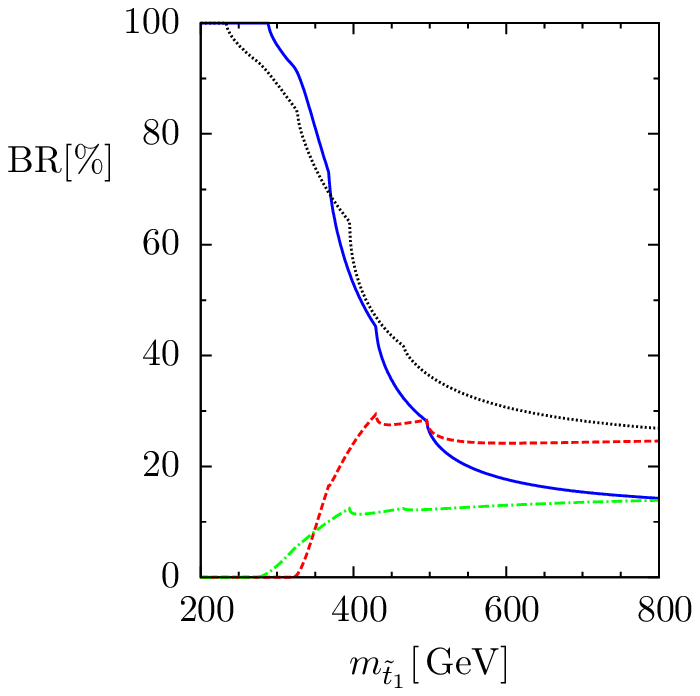}}
\put(1,0.15){(c)}
\put(9,0.15){(d)}
\end{picture}

\caption{\it The CP asymmetry ${\cal A}$, see~(\ref{Anumerics}),  for 
  stop  pair production 
  $q \bar q \to \tilde t_\alpha(\bar s)\,\tilde t_\alpha^\ast(\bar s)$
  and  decay  $\tilde  t_\alpha  \to  b  \,  \tilde\chi^+_1$,  $\tilde
  t_\alpha^\ast   \to  \bar   b   \,\tilde\chi^-_2$,  at   $\sqrt{\bar
  s\,}=(m_{\tilde   t_1}  +  m_{\tilde   t_2})/2$,  as   functions  of
  $\phi_{A_t}$  and $m_{\tilde  t_1}$.  Also  shown are  the branching
  ratios ${\rm BR}(\tilde t_1 \to b \,\tilde\chi^+_1 )$ (solid, blue),
  ${\rm BR}(\tilde t_1 \to  b \,\tilde\chi^+_2 )$ (dashed, red), ${\rm
  BR}(\tilde  t_2 \to  b  \,\tilde\chi^+_1 )$  (dotted, black),  ${\rm
  BR}(\tilde t_2  \to b  \,\tilde\chi^+_2 )$ (dash-dotted,  green), as
  functions of  $\phi_{A_t}$ and  $m_{\tilde t_1}$. In  the $m_{\tilde
  t_1}$-plots  (c) and (d), $M_{\tilde  Q} = M_{\tilde  U} = M_{\tilde
  D}$ is varied,  whilst the other MSSM parameters  are fixed as given
  in the text of subsection~\ref{sec:stopCP}.}
\label{fig:AsymmandBR}
\end{figure}

\begin{figure}[t]
\centering
\begin{picture}(16,7.)
\put(-2.15,-13.9){\includegraphics{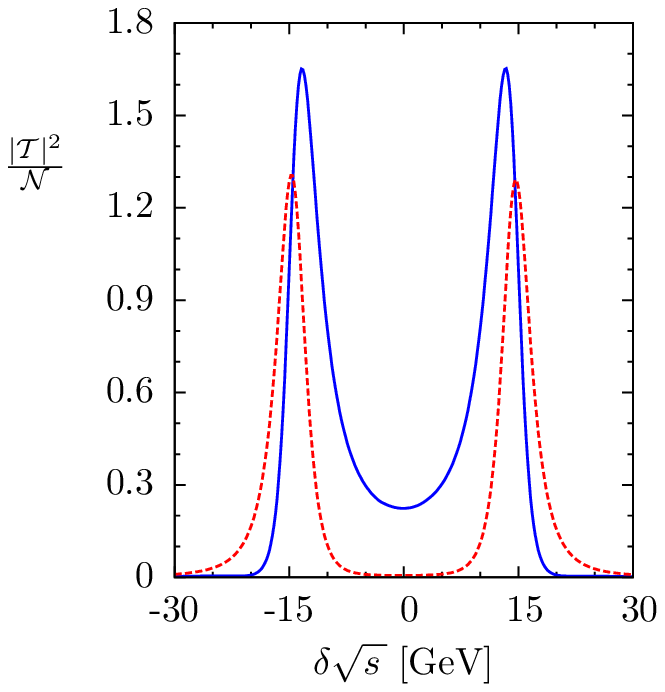}}
\put(6,-13.9){\includegraphics{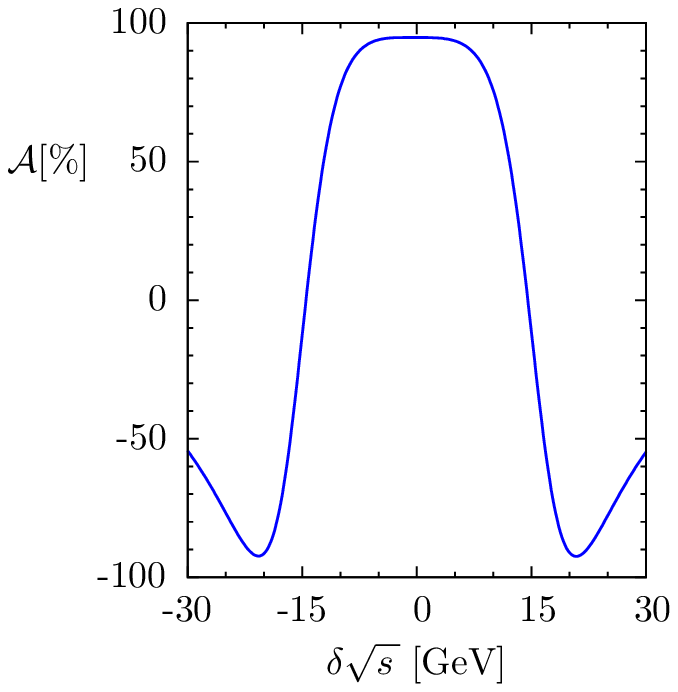}}
\put(1,0.15){(a)}
\put(9,0.15){(b)}
\end{picture}
\caption{\it 
       The dependence on $\delta\sqrt s = (\sqrt{s_1} - \sqrt{s_2})/2$
       of (a)~the amplitude squared $|{\mathcal T}(s_1,s_2)|^2$~(\ref{ampli12a})
       (up to an arbitrary normalization $\mathcal N$), with
       $\sqrt{ s_{1,2}} =\sqrt{\bar s \,} \pm \delta\sqrt s$ 
       at the resonance region $\sqrt{\bar s\,}=(m_{\tilde t_1}+m_{\tilde    
       t_2})/2$, for stop pair production 
       $q \bar q \to \tilde t_\alpha(s_1)\,\tilde t_\alpha^\ast(s_2)$
       and decay $\tilde t_\alpha \to b \, \tilde\chi^+_1$,
       $\tilde t_\alpha^\ast \to \bar b \, \tilde\chi^-_2$ (solid, blue),
       and the CP-conjugate amplitude squared~(\ref{ampli12b}) for the 
       process $\tilde t_\alpha \to b \, \tilde\chi^+_2$,
       $\tilde t_\alpha^\ast \to \bar b \, \tilde\chi^-_1$ (dashed, red),
       and (b)~the corresponding CP asymmetry~(\ref{Anumerics}),
       with MSSM parameters as given in the text
       of subsection~\ref{sec:stopCP}.
 }
\label{fig:asymmABSqrts}
\end{figure}

For our  numerical example, we  choose the MSSM  parameters $M_{\tilde
  Q}=  M_{\tilde U}= M_{\tilde  D}=470$~GeV, $M_{\tilde  L}= M_{\tilde
  E}=120$~GeV, $|A_t|  = |A_b| = |A_\tau|  = 110$~GeV, $M_2 =  2 M_1 =
|\mu|=250$~GeV, $m_{\tilde g}=710$~GeV,  $\tan\beta=5$, and the phases
$\phi_{A_t}=                                                    \pi/4$,
$\phi_{A_b}=\phi_{A_\tau}=\phi_{M_1}=\phi_{\mu}=\phi_{\tilde     g}=0$.
At tree level, the resulting stop masses, widths, and mixing angle are
$m_{\tilde t_{ \{ 1,2  \} }}=\{ 485,514\} $~GeV, $\Gamma_{\tilde t_{\{
    1,2  \}}}= \{5.3,9.1\}$~GeV, $\cos\theta_{\tilde  t}=-0.725$.  The
gaugino  masses are $m_{\tilde  \chi_{\{1,2\}}^\pm}= \{187,321\}$~GeV,
$m_{\tilde \chi_{ \{1,2,3,4\}}^0}=\{115,194,256,322\} $~GeV.

As  an illustrative  example, we  study stop  pair production  via the
process $q  \bar q \to \tilde  t_\alpha\tilde t_\alpha^\ast$, decaying
into  charginos: $\tilde t_\alpha  \to b  \tilde\chi^+_1~(A)$, $\tilde
t_\alpha^\ast \to \bar b \tilde\chi^-_2~(B)$.  At the resonant region,
$\sqrt{\bar  s\,}=(m_{\tilde t_1}+m_{\tilde  t_2})/2$,  the transition
amplitude     squared    and     its    CP-conjugate     one    follow
from~(\ref{eq:respart}) to be
\begin{eqnarray}
|{\mathcal T}(\bar s)|^2 &= &
{\rm Tr}\Big[T_A(\bar s)\,\left(\Delta(\bar s)\right)^2\, 
             T_B(\bar s)\,\left(\Delta^{\dagger }(\bar s)\right)^2\Big]\;,\\[2mm]
|\overline{\mathcal T}(\bar s)|^2 &= &
{\rm Tr}\Big[T_B(\bar s)\,\left(\Delta(\bar s)\right)^2\, 
             T_A(\bar s)\,\left(\Delta^{\dagger }(\bar s)\right)^2\Big]\;.
\end{eqnarray}
In Fig.~\ref{fig:AsymmandBR},  we show the phase  $\phi_{A_t}$ and the
stop mass dependence of the CP asymmetry
\begin{eqnarray}
{\mathcal A} &= &
       \frac{|{\mathcal T} |^2 - |\overline {\mathcal T} |^2}
            {|{\mathcal T} |^2 + |\overline {\mathcal T} |^2} \; .
\label{Anumerics}
\end{eqnarray}
We observe that the CP asymmetry can reach values of order one, thanks
to a  maximal mixing of the  two stop states in  production and decay.
The  CP asymmetry  would decrease  as stop  mass  splitting increases,
which can occur  for large mass differences $M_{\tilde  Q} - M_{\tilde
  U}$.

In Fig.~\ref{fig:asymmABSqrts}, we study in more detail the dependence
of the amplitudes [cf.~(\ref{2to4scatamplitude})],
\begin{eqnarray}
|{\mathcal T}(s_1,s_2)|^2 &= &
 {\rm Tr}\Big[\, T_A(s_1)\,\Delta(s_1)\,\Delta(s_2)\,
                 T_B(s_2)\,\Delta^\dagger(s_2)\,\Delta^\dagger(s_1)  \,
          \Big]\; ,\label{ampli12a}\\[2mm]
|\overline{\mathcal T}(s_1,s_2)|^2 &= &
 {\rm Tr}\Big[\, T_B(s_1)\,\Delta(s_1)\,\Delta(s_2)\,
                 T_A(s_2)\,\Delta^\dagger(s_2)\,\Delta^\dagger(s_1)  \,
          \Big]\; ,
\label{ampli12b}
\end{eqnarray}
and the corresponding CP asymmetry as defined in~(\ref{Anumerics}), as
functions  of $\delta\sqrt s  = (\sqrt{s_1}  - \sqrt{s_2})/2$.  In our
numerical estimates, we use the parameterization
\begin{eqnarray}
  \sqrt{ s_1} =\sqrt{\bar s \,} + \delta\sqrt s\; , &&
  \sqrt{ s_2} =\sqrt{\bar s \,} - \delta\sqrt s\; ,
\end{eqnarray}
around   the    resonance   region   $\sqrt{\bar    s\,}   =(m_{\tilde
  t_1}+m_{\tilde     t_2})/2$.      As     can    be     seen     from
Fig.~\ref{fig:asymmABSqrts},  the CP  asymmetry ${\cal  A}$ can  be of
order  1 for the  experimentally testable  interval of  $\delta\sqrt s
\sim 20$~GeV.

\begin{figure}[t]
\centering
\begin{picture}(16,7.5)
\put(-2.15,-13.9){\includegraphics{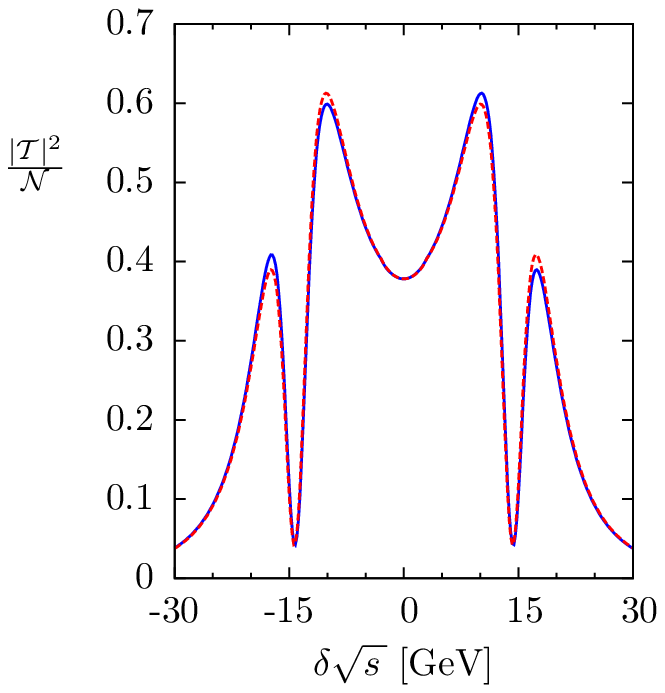}}
\put(6,-13.9){\includegraphics{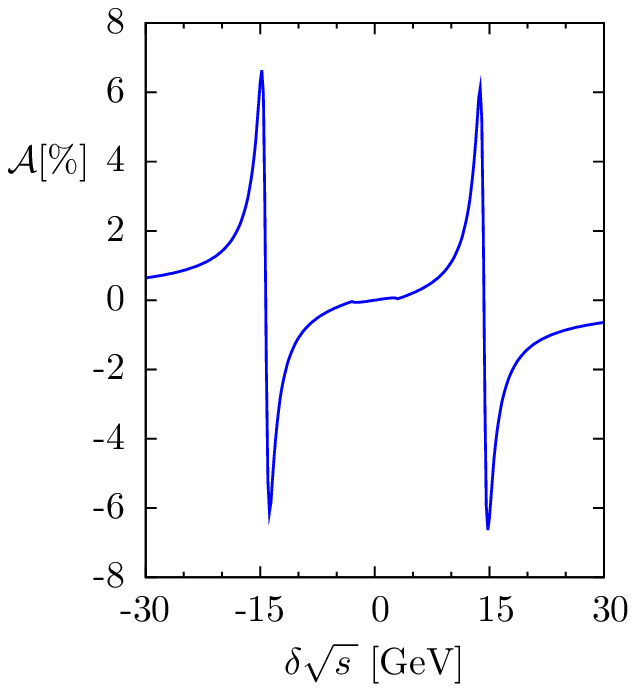}}
\put(1,0.15){(a)}
\put(9,0.15){(b)}
\end{picture}
\caption{\it The  same as  in Fig.~\ref{fig:asymmABSqrts} for  (a) the
  stop  pair  production  $q  \bar q  \to  \tilde  t_\alpha(s_1)\, \tilde
  t_\alpha^\ast(s_2)$    and   decay    $\tilde    t_\alpha   \to    b \,
  \tilde\chi^+_1$, $\tilde  t_\alpha^\ast \to \bar  b \, \tilde\chi^-_1$
  (solid, blue),
  and  the  CP-conjugate amplitude  squared  for  the process  $\tilde
  t_\alpha  \to b \, \tilde\chi^+_1$, $\tilde  t_\alpha^\ast \to  \bar b \,
  \tilde\chi^-_1$ (dashed, red), and for (b) the corresponding CP asymmetry. }
\label{fig:asymmABSqrts2}
\end{figure}

Finally,   in   Fig.~\ref{fig:asymmABSqrts2}   we  display   numerical
estimates of stop decays into  the {\em same} states: $\tilde t_\alpha
\to   b  \tilde\chi^+_1~(A)$,   $\tilde  t_\alpha^\ast   \to   \bar  b
\tilde\chi^-_1~(\bar A)$.   The corresponding CP  asymmetry ${\cal A}$
is  non-vanishing,   of  order  5\%,  only  for   non-zero  values  of
$\delta\sqrt s$. Hence, CP violation in the same channel would be more
difficult to probe for the stop scenario of the MSSM under study.

Before  closing this  section, it  is  interesting to  comment on  the
discovery potential of  the LHC in measuring a  non-zero CP asymmetry.
As shown  in~\cite{Deppisch:2010nc}, the minimal luminosity~${\mathcal
  L}$ to  observe a non-vanishing  CP asymmetry~${\mathcal A}$  may be
estimated by
\begin{eqnarray}
{\mathcal L} &=&  \frac{n^2}{\sigma}\left( \frac{1}{{\mathcal A^2}} -1
\right),
\label{eq:lumi}
\end{eqnarray}
where $n$ is the number  of standard deviations (confidence level) and
$\sigma$ is  the cross  section of the  signal events.   Obviously, to
obtain realistic values  of statistical significances and luminosities
requires  detailed Monte--Carlo  simulations.  Since  such a  study is
beyond the scope  of the present paper, here we only  offer an initial
estimate.

For our benchmark  scenario, with $m_{\tilde t_{1,2}}\approx 500$~GeV,
the cross section for stop pair production is $\sigma_p(p p \to \tilde
t_\alpha      \tilde     t_\alpha)\approx\mathcal{O}(700)$~fb,     for
$\alpha=1,2$, at  the next-to-leading  order, according to  the public
code  \texttt{PROSPINO}~\cite{Beenakker:1997ut}.   The stop  branching
ratios into  charginos are found approximately to  be ${\rm BR}(\tilde
t_\alpha \to b \,\tilde\chi^\pm_j )\approx 20\%$, for each $j=1,2$ and
$\alpha=1,2$  (see panels (b)  and (d)  in Fig.~\ref{fig:AsymmandBR}).
Correspondingly, the  leptonic ($\ell=e,\mu$) branching  ratios of the
charginos        are       ${\rm        BR}(\tilde\chi^\pm_j       \to
\tilde\nu_\ell\,\ell)\approx  30\%$,   using   the   formulae   quoted
in~\cite{Kittel:2004rp}      and      the      fact      that      the
sneutrinos~$\tilde\nu_\ell$   decay   invisibly   in   our   scenario,
i.e.~$\tilde\nu_\ell\to \nu_\ell\,\tilde\chi^0_1$.

Taking  all the above  estimates into  account, we  find a  net signal
cross  section   $\sigma\approx  \mathcal{O}(3)$~fb,  for   stop  pair
production and  decay.  In addition  to using standard cuts  to reduce
the background and  isolate the signal, one has  still to identify the
charges  of  the  final   leptons  and  the  mother  chargino  states,
$\tilde\chi^\pm_1$ or $\tilde\chi^\pm_2$,  for a proper measurement of
the  CP  asymmetries.  Recently,  a  similar  study of  triple-product
asymmetries         in         stop        pair-production         and
decay~\cite{MoortgatPick:2010wp}   has  shown   that  the   method  of
kinematic event  selection allows such an identification  of the decay
chains,  if  the particle  masses  are  known,  presumably from  other
measurements.  Because  of the kinematic cuts and  the event selection
criteria,  the signal  events are  reduced  typically by  an order  of
magnitude, with the surviving CP-even background being similar in size
to the signal~\cite{MoortgatPick:2010wp}.  Based on these results, our
CP-asymmetry is expected to be reduced  by half, such that we are left
with ${\mathcal A}\approx 50\%$  in the best case scenario.  Inserting
these numbers  into Eq.~(\ref{eq:lumi}), we  find ${\mathcal L}\approx
n^2~10~\rm{fb}^{-1}$.    Thus,  LHC  luminosities   of  order   a  few
$10~\rm{fb}^{-1}$  would   be  sufficient  to   observe  resonant  EPR
CP asymmetries of order one.

\newpage

\section{Conclusions}\label{sec:concl}

We have analyzed resonant CP-violating Einstein--Podolsky--Rosen~(EPR)
correlations  in the  pair  production and  decay  of unstable  scalar
particles  at  high-energy  colliders.    We  have  shown  that  as  a
consequence of unitarity and CPT invariance of the ${\rm S}$-matrix, a
minimum number of linearly  independent decay matrices associated with
the  unstable  scalar  states  are  necessary in  order  to  obtain  a
non-vanishing result  for CP-odd observables  that are also  odd under
C-conjugation,  but   even  under  P   transformations.   For  $2\to2$
scatterings,  at  least  three  such independent  decay  matrices  are
needed, whereas  for $2\to4$  scatterings, only two  independent decay
matrices are required.   Even though our study involved  the mixing of
scalar particles only, we may  safely conjecture that our results will
also hold true for the mixing of unstable fermions and vector bosons.

As  a direct  application  of  the above  theorem,  we have  presented
numerical estimates  of CP  asymmetries for the  correlated production
and decay of supersymmetric scalar top--anti-top pairs at the LHC.  We
have  explicitly  demonstrated  that   the  CP  asymmetries  could  be
sizeable, reaching  values of  order one and  so making  them directly
testable.   Our~analysis  also  led  us  to the  development  a  novel
spinorial  trace formalism,  which helped  us to  efficiently evaluate
lengthy expressions  of squared amplitudes  describing resonant scalar
transitions.  The formalism presented  here could be extended to other
unstable-particle  systems,  predicted in  minimal  extensions of  the
Standard     Model    that     include    mixing     of    Higgs-boson
states~\cite{resHiggs,resHiggs2}  or  the  mixing  of  heavy  unstable
neutrinos~\cite{resNeutrino}.     It~is    interesting    to   study
systematically  possible   phenomena  of  resonant   CP-violating  EPR
correlations in these systems both  for the LHC and future high-energy
colliders.


\subsection*{Acknowledgments}
We thank  Jose~Bernab\'eu, Peter~Millington and Federico~v.~d.~Pahlen
for useful discussions.  OK~gratefully acknowledges partial support by
MICINN project FPA.2006-05294 and CPAN, and AP the STFC research grant,
ref: ST/J000418/1.

\bigskip

\newpage

\begin{appendix}
\renewcommand{\thesubsection}{\Alph{section}.\arabic{subsection}}
\renewcommand{\theequation}{\Alph{section}.\arabic{equation}}
\setcounter{equation}{0}

\section{Spinorial Trace Technique for Two Scalar Mixing}
\label{sec:Matrixformalism}

Here we present our conventions  for our spinorial trace technique, as
well as useful  identities and analytical expressions for  the case of
the two-unstable-particle mixing.

\subsection{Preliminaries}

A  basic   element  of  our   formalism  are  the   generalized  Pauli
matrices~\cite{mssm}:
\begin{eqnarray}
 \sigma^{\mu} &=& (\sigma^0, \mbox{\boldmath$ \sigma$ } )\; ,
\qquad
 \bar\sigma^{\mu}\; =\; (\sigma^0, -\mbox{\boldmath$ \sigma$ } )\; ,
\label{genPauli}
\end{eqnarray}
with
\begin{equation}
\begin{array}{cccccccc}
\sigma^{0}  = 
\left(\begin{array}{cc}
1 & 0\\
0 & 1 
\end{array}\right), & 
\sigma^{1}  = 
\left(\begin{array}{cc}
0 & 1\\
1 & 0 
\end{array}\right),  & 
\sigma^{2} = 
\left(\begin{array}{cc}
0 & -i\\
i & 0
\end{array}\right), &  
\sigma^{3}  = 
\left(\begin{array}{cc}
1 & 0\\
0 & -1
\end{array}\right). 
\end{array}
\label{Pauli}
\end{equation}
Note   that   these   are   also   the   generators   of   the   ${\rm
  SL}(2,\mathbb{C})$ group.  Then, any $2\times 2$-dimensional complex
matrix $A$ can be expanded as
\begin{equation}
\label{en}
A\ =\ A_\mu \sigma^{\mu}
\ =\  A^\mu \bar\sigma_{\mu}\; , \qquad
A^{-1} \ =\ \frac{ A_\mu \bar\sigma^{\mu}}{A_\nu A^\nu}\ ,
\end{equation}
where $A_\nu A^\nu  = {\rm det} A$ and  the complex components $A^\nu$
may be evaluated as
\begin{equation}
  \label{ex}
A^\nu \ = \ \frac{1}{2}{\rm Tr}\left[ A\, \bar\sigma^{\nu} \right]\; .
\end{equation}
In calculating traces, the following relations are useful~\cite{mssm}:
\begin{eqnarray}
\sigma^{\mu} \bar \sigma^{\nu} \sigma^{\rho} +
\sigma^{\rho} \bar \sigma^{\nu} \sigma^{\mu}
&=&
2 ( g^{\mu\nu}\sigma^{\rho} - 
    g^{\mu\rho}\sigma^{\nu} +  
    g^{\nu\rho}\sigma^{\mu}  )\; , 
\\[2mm]
\sigma^{\mu} \bar \sigma^{\nu} \sigma^{\rho} -
\sigma^{\rho} \bar \sigma^{\nu} \sigma^{\mu}
&=&
- 2 i \varepsilon^{\mu\nu\rho\lambda}\sigma^{\lambda}\; ,
\\[2mm]
\bar\sigma^{\mu}  \sigma^{\nu} \bar\sigma^{\rho} +
\bar\sigma^{\rho} \sigma^{\nu} \bar\sigma^{\mu}
&=&
2 ( g^{\mu\nu}\bar\sigma^{\rho} - 
    g^{\mu\rho}\bar\sigma^{\nu}+  
    g^{\nu\rho}\bar\sigma^{\mu}  )\; ,
\\[2mm]
\bar\sigma^{\mu}  \sigma^{\nu} \bar\sigma^{\rho} -
\bar\sigma^{\rho} \sigma^{\nu} \bar\sigma^{\mu}
&=&
 2 i \varepsilon^{\mu\nu\rho\lambda} \bar\sigma^{\lambda}\; ,
\end{eqnarray}
with the convention  $\varepsilon_{0123}=+1$.  For instance, the traces
involving two and four generalized Pauli matrices are given by
\begin{eqnarray}
{\rm Tr}\left[ \sigma^{\mu}\bar\sigma^{\nu} \right] &=& 2 g^{\mu\nu}\; ,
\\[2mm]
{\rm Tr}\left[
  \sigma^{\mu}\bar\sigma^{\nu}\sigma^{\rho}\bar\sigma^{\lambda} \right] &=&
2\, \left(
g^{\mu\nu}g^{\rho\lambda} - 
    g^{\mu\rho}g^{\nu\lambda} +  
    g^{\nu\rho}g^{\mu\lambda}
-i \varepsilon^{\mu\nu\rho\lambda} \right)\; .
\label{4sigmatrace}
\end{eqnarray}

\subsection{Expansion of the Scalar Propagator Matrix}
\label{sec:exPropagatormatrix}

We  can now  use  the  4-dimensional basis  of  the generalized  Pauli
matrices    $\sigma^\mu$   to    expand    the   inverse    propagator
in~(\ref{invProp}), i.e.
\begin{eqnarray}
\Delta^{-1}(s) &=& 
\left(
 \begin{array}{cc}
                    s - M_1^2  &  - {\rm Im}\Pi_{12}^{\rm abs}(s)   \\[2mm]
   {\rm Im}\Pi_{12}^{\rm abs}(s)  &        s - M_2^2
 \end{array} \right)
 +i \, \left(
 \begin{array}{cc}
 \Pi_{11}^{\rm abs}(s)            &    {\rm Re}\Pi_{12}^{\rm abs}(s)   \\[2mm]
  {\rm Re}\Pi_{12}^{\rm abs}(s)   &    \Pi_{22}^{\rm abs}(s)
 \end{array} \right)\; . \quad
\end{eqnarray}
To this  end, we  use~(\ref{en}) to write  the $2\times  2$ propagator
matrix $\Delta (s)$ as
\begin{equation}
\Delta (s) \ = \ 
\frac{ \Delta^{-1}_\mu (s)\, \bar\sigma^{\mu}}
                {{\rm det}\left[\Delta^{-1}(s)\right]}\ ,
\end{equation}
where the complex 4-vector components $\Delta^{-1}_\mu (s)$ are given by
\begin{eqnarray}
  \label{Dparts}
\Delta_{0}^{-1}(s) &=& s\ -\
 \frac{1}{2}\: (M_1^2+ M_2^2 )\
   \ +\ \frac{i}{2}\: \Big[\Pi_{11}^{\rm abs}(s) 
+ \Pi_{22}^{\rm abs}(s) \Big]\; ,
\\[2mm]
\Delta_{1}^{-1}(s) &=& \phantom{+} i \, {\rm Re}\, \Pi_{12}^{\rm abs}(s)\; ,
\\[2mm]
\Delta_{2}^{-1}(s) &=& -i \,{\rm Im}\, \Pi_{12}^{\rm abs}(s)\; ,
\\[2mm]
\Delta_{3}^{-1}(s) &=&
            \frac{1}{2}\: (M_2^2- M_1^2 )\ +\ 
            \frac{i}{2}\: \Big[\Pi_{11}^{\rm abs}(s)\ -\ 
\Pi_{22}^{\rm abs}(s) \Big]\; ,
\end{eqnarray}
and
\begin{eqnarray}
{\rm det}\left[\Delta^{-1}(s)\right] &=&
(s-M_1^2)\,(s-M_2^2)\ +\ |\Pi_{12}^{\rm abs}(s)|^2\
-\ \Pi_{11}^{\rm abs}(s)\: \Pi_{22}^{\rm abs}(s)\nonumber\\[2mm]  &&
  +\ i\,\Big[\, \Pi_{11}^{\rm abs}(s)\:(s-M_2^2) \
+\ \Pi_{22}^{\rm abs}(s)\:(s-M_1^2)\,
    \Big]\; .
\end{eqnarray}
Correspondingly, the Hermitian decay matrices $T_{A,B,C}$
for the final states ${A=(f_k\,\tilde{f}_{\tilde {l} })}$, 
see~(\ref{Tab}), etc., may be each decomposed 
as $T(s) = T_\mu(s)\: \sigma^\mu$, with the real coefficients
\begin{eqnarray}
  \label{Thparts}
T_{0}(s) &=&
\frac{1}{2} \Big[T^{11}(s) +T^{22}(s) \Big]
\; ,\\[3mm]
T_{1}(s) &=& \;
\phantom{+}{\rm Re}\,T^{12}(s) 
%
\; ,\\[3mm]
T_{2}(s) &=&    \;
      -{\rm Im}\,T^{12}(s)
%
\; ,\\[3mm]
T_{3}(s) &=&
\frac{1}{2}\Big[T^{11}(s) -T^{22}(s) \Big]
%
\; .
\end{eqnarray}
where
\begin{eqnarray}
T^{12}(s)    &=&
  \left(C^{1,L\ast}_{k\tilde l}C^{2,L}_{k\tilde l}  +
        C^{1,R\ast}_{k\tilde l}C^{2,R}_{k\tilde l} \right)  
        (s -m_{k}^2- m_{\tilde l}^2 )
  - 2\left(C^{1,R\ast}_{k\tilde l}C^{2,L}_{k\tilde l}  +
       C^{1,L\ast}_{k\tilde l}C^{2,R}_{k\tilde l} 
 \right)m_{k}m_{\tilde l}\; , \nonumber
\\[3mm]
T^{\alpha\alpha}(s)    &=&
  \left(|C^{\alpha,L\ast}_{k\tilde l}|^2  +
        |C^{\alpha,R\ast}_{k\tilde l}|^2 \right)  
        (s -m_{k}^2- m_{\tilde l}^2 )
- 4 \,{\rm Re}\left(C^{\alpha,L}_{k\tilde l} C^{\alpha,R\ast}_{k\tilde l} 
 \right)m_{k}m_{\tilde l}\; . 
\end{eqnarray}
Note that the partial scalar decay widths at tree level are
\begin{eqnarray}
\Gamma_\alpha\left(\widetilde S_\alpha \to f_k \tilde f_{\tilde l}\right)
&=&
  \frac{\lambda^{1/2} (M^2_{\alpha},m^2_k,m^2_{\tilde l}) }
{16 \, \pi \, M^3_{\alpha}} \;
T^{\alpha\alpha}(M^2_{\alpha})\; .
\end{eqnarray}

\newpage

\end{appendix}



\begin{thebibliography}{99}

\bibitem{EPR} A. Einstein, B. Podolsky and N. Rosen, Phys.\ Rev.\ {\bf
  47} (1935) 777.

\bibitem{Bell} J. S. Bell, ``Speakable and Unspeakable in Quantum
  Mechanics,'' Cambridge University Press, 1987.

\bibitem{Bernabeu:2003ym}
  J.~Bernabeu, N.~E.~Mavromatos and J.~Papavassiliou,
  Phys.\ Rev.\ Lett.\  {\bf 92} (2004) 131601
  [arXiv:hep-ph/0310180].

\bibitem{resScalar}
%
  A.~Pilaftsis,
  Nucl.\ Phys.\  B {\bf 504}, 61 (1997)
  [arXiv:hep-ph/9702393].

\bibitem{stauCP}  
S.~Y.~Choi and  M.~Drees,
  Phys.\ Lett.\ B\ {\bf 435} (1998) 356
  [arXiv:hep-ph/9805474];\\
S.~Y.~Choi, M.~Drees, B.~Gaissmaier and J.~S.~Lee,
  Phys.\ Rev.\ D\ {\bf 64} (2001) 095009
  [arXiv:hep-ph/0103284].


\bibitem{mssm}
  H.~E.~Haber and G.~L.~Kane,
  Phys.\ Rept.\  {\bf 117}, 75 (1985);\\
%
  H.~P.~Nilles,
  Phys.\ Rept.\  {\bf 110} (1984) 1;\\
%
  M.~Drees, R.~Godbole and P.~Roy,
  \emph{Theory and phenomenology of sparticles}, 
  World Scientific, Singapore (2004).
%

\bibitem{Heinemeyer:2010mm}
  S.~Heinemeyer, H.~Rzehak and C.~Schappacher,
  Phys.\ Rev.\  D {\bf 82} (2010) 075010
  [arXiv:1007.0689 [hep-ph]].


\bibitem{Hlucha:2011yk}
  H.~Hlucha, H.~Eberl and W.~Frisch,
  arXiv:1104.2151 [hep-ph].




%
%


\bibitem{Bartl:2003he}
  A.~Bartl, S.~Hesselbach, K.~Hidaka, T.~Kernreiter and W.~Porod,
  Phys.\ Rev.\  D {\bf 70}, 035003 (2004)
  [arXiv:hep-ph/0311338].


\bibitem{Eberl:2009xe}
  H.~Eberl, S.~M.~R.~Frank and W.~Majerotto,
  Eur.\ Phys.\ J.\  C {\bf 70}, 1017 (2010)
  [arXiv:0912.4675 [hep-ph]].




\bibitem{CutRules}
R.~E.~Cutkosky, J.\  Math.\  Phys. {\bf 1}, 429 (1960);\\
%
  M.~J.~G.~Veltman,
  ``Diagrammatica: The Path to Feynman rules,''
{\it  Cambridge, UK: Univ. Pr. (1994) (Cambridge lecture notes in
  physics, 4)}.
%

\bibitem{Marek} A.~Pilaftsis and M.~Nowakowski,
  Phys.\ Lett.\  B {\bf 245} (1990) 185.

\bibitem{nima}  N.~Arkani-Hamed, J.~L.~Feng, L.~J.~Hall and H.~C.~Cheng,
  Nucl.\ Phys.\  B {\bf 505} (1997)~3
  [arXiv:hep-ph/9704205];\\
D.~Bowser-Chao and W.~Y.~Keung,
  Phys.\ Rev.\  D {\bf 56} (1997) 3924
  [arXiv:hep-ph/9704219].


\bibitem{reviewPT}
  D.~Binosi and J.~Papavassiliou,
  Phys.\ Rept.\  {\bf 479} (2009) 1
  [arXiv:0909.2536 [hep-ph]].

\bibitem{Deppisch:2010nc}
  F.~F.~Deppisch and O.~Kittel,
  JHEP {\bf 1006}, 067 (2010)
  [arXiv:1003.5186 [hep-ph]];
%
  JHEP {\bf 0909}, 110 (2009)
  [Erratum-ibid.\  {\bf 1003}, 091 (2010)]
  [arXiv:0905.3088 [hep-ph]].


\bibitem{Beenakker:1997ut}
  W.~Beenakker, M.~Kramer, T.~Plehn, M.~Spira and P.~M.~Zerwas,
  Nucl.\ Phys.\  B {\bf 515}, 3 (1998)
  [arXiv:hep-ph/9710451].

\bibitem{Kittel:2004rp}
  O.~Kittel,
  arXiv:hep-ph/0504183.

\bibitem{MoortgatPick:2010wp}
  G.~Moortgat-Pick, K.~Rolbiecki and J.~Tattersall,
  Phys.\ Rev.\  D {\bf 83}, 115012 (2011)
  [arXiv:1008.2206 [hep-ph]];
%
  G.~Moortgat-Pick, K.~Rolbiecki, J.~Tattersall and P.~Wienemann,
  JHEP {\bf 1001}, 004 (2010)
  [arXiv:0908.2631 [hep-ph]].

\bibitem{resHiggs}
  J.~R.~Ellis, J.~S.~Lee and A.~Pilaftsis,
  Phys.\ Rev.\  D {\bf 70} (2004) 075010
  [arXiv:hep-ph/0404167];
  Phys.\ Rev.\  D {\bf 72} (2005) 095006
  [arXiv:hep-ph/0507046];
  Phys.\ Rev.\  D {\bf 71} (2005) 075007
  [arXiv:hep-ph/0502251].

\bibitem{resHiggs2} 
  J.~Bernabeu, D.~Binosi and J.~Papavassiliou,
  JHEP {\bf 0609} (2006) 023
  [arXiv:hep-ph/0604046];\\
%
  H.~K.~Dreiner, O.~Kittel and F.~von der Pahlen,
  JHEP {\bf 0801}, 017 (2008)
  [arXiv:0711.2253 [hep-ph]];\\
%
  O.~Kittel and F.~von der Pahlen,
  JHEP {\bf 0808}, 030 (2008)
  [arXiv:0806.4534 [hep-ph]].


\bibitem{resNeutrino}
  S.~Bray, J.~S.~Lee and A.~Pilaftsis,
  Nucl.\ Phys.\  B {\bf 786} (2007) 95
  [arXiv:hep-ph/0702294];\\
  S.~Blanchet, Z.~Chacko, S.~S.~Granor and R.~N.~Mohapatra,
  Phys.\ Rev.\  D {\bf 82} (2010) 076008
  [arXiv:0904.2174 [hep-ph]].



\end{thebibliography}
\end{document}